\pdfoutput=1
\documentclass[aps,
prd,
reprint,
superscriptaddress,
amsmath,amssymb,
]{revtex4-1}
\usepackage[T1]{fontenc}
\usepackage{ae,aecompl}
\usepackage[english]{babel}
\usepackage{graphicx}
\usepackage{dcolumn}
\usepackage{bm}
\usepackage{mathrsfs} 
\usepackage{stmaryrd} 
\usepackage{multirow} 
\usepackage{hyperref}
\usepackage{color} 
\usepackage{mathtools} 
\usepackage{enumitem} 
\usepackage{mathtools} 

\usepackage{placeins} 
\usepackage{soul} 

\newcommand{\aap}{{Astron. Astrophys. }}

\definecolor{aqua}{rgb}{0.0, 1.0, 1.0}

\begin{document}

\preprint{APS/123-PRD}

\title{Gaussian regression and power spectral density estimation with missing data: The MICROSCOPE space mission as a case study}

\author{Quentin Baghi}
\email{quentin.baghi@onera.fr}
\affiliation{
ONERA - The French Aerospace Lab,
29 avenue de la Division Leclerc, 92320 Ch\^{a}tillon, France
}
\author{Gilles M\'{e}tris}%
\email{gilles.metris@oca.eu}
\affiliation{
Universit\'{e} Nice Sophia Antipolis, CNRS, Observatoire de la C\^{o}te d'Azur, G\'{e}oazur UMR 7329,
250 rue Albert Einstein, Sophia Antipolis, 06560 Valbonne, France
}
\author{Jo\"{e}l Berg\'{e}}
\author{Bruno Christophe}
\author{Pierre Touboul}
\author{Manuel Rodrigues}
\affiliation{
ONERA - The French Aerospace Lab,
29 avenue de la Division Leclerc, 92320 Ch\^{a}tillon, France
}

\date{\today}

\begin{abstract}
We present a Gaussian regression method for time series with missing data and stationary residuals of unknown power spectral density (PSD). The missing data are efficiently estimated by their conditional expectation as in universal Kriging, based on the circulant approximation of the complete data covariance. After initialization with an autoregessive fit of the noise, a few iterations of estimation/reconstruction steps are performed until convergence of the regression and PSD estimates, in a way similar to the expectation-conditional-maximization  algorithm. The estimation can be performed for an arbitrary PSD provided that it is sufficiently smooth. The algorithm is developed in the framework of the MICROSCOPE space mission whose goal is to test the weak equivalence principle (WEP) with a precision of $10^{-15}$.
We show by numerical simulations that the developed method allows us to meet three major requirements: to maintain the targeted precision of the WEP test in spite of the loss of data, to calculate a reliable estimate of this precision and of the noise level, and finally to provide consistent and faithful reconstructed data to the scientific community.

\begin{description}
\item[PACS numbers]
\end{description}
\end{abstract}

\maketitle

\section{\label{sec:intro}Introduction}

Linear regression and spectral estimation in time series are involved in the analysis of various experimental measurements in physics. However the data sets necessary to extract relevant scientific information may be partially unavailable or may suffer from short interruptions. This is likely to arise in astrophysical measurements such as asteroseismology \cite{Garcia} or experiments in fundamental physics where long integration times are needed. Relevant examples are ultra-sensitive tests of general relativity such as MICROSCOPE \cite{Touboul} or forthcoming eLISA \cite{eLISA} and its technological demonstrator LISA Pathfinder \cite{LISAPF} where periodic interruptions may be expected \cite{Cutler}. In each case, when testing the weak equivalence principle (WEP) or paving the way for space-based gravitational waves detectors, the aim is to discover faint signals buried in a correlated noise, but also to characterize the noise level, which requires a careful treatment of missing data.

In the MICROSCOPE space mission the measurement is performed by electrostatic accelerometers with a high sensitivity, on-board a microsatellite orbiting Earth in a quasi-circular orbit. The measurement is therefore likely to be perturbed by unpredictable events occurring in the satellite equipment. Among them are (in descending order of probability) crackles of the cold gas tank walls due to depressurization, crackles of the multilayer insulation (MLI) coating due to temperature changes in flight and micrometeorite impacts. These phenomena are expected to trigger short acceleration peaks that may be above the saturation limit of the accelerometers. Another but less likely source of interruptions are telemetry losses.
As a result, during the corresponding time spans there will be a corrupted, or an absence of, information that we call ``missing data'' or ``data gaps''.

Data gaps can have a significant impact on the uncertainty of the ordinary least-squares regression. In the case of MICROSCOPE this uncertainty can be increased by more than one order of magnitude for a WEP test session \cite{Baghi}, therefore an adapted data processing is necessary. 

The general problem under study in this paper is the uni-variate linear regression problem with missing data and stationary Gaussian noise, which we define as follows. Assume that we measure a unique realization of a large random vector $y$ with a regular sampling, where some points are missing. This vector is assumed to have a Gaussian distribution with mean $\mu$ and covariance $\Sigma$. On the one hand, the mean is a linear combination of known signals whose amplitudes $\beta_i$ are to be determined. These amplitudes are referred to as ``regression parameters''. On the other hand, the covariance is assumed to describe a stationary process, hence $\Sigma$ is Toeplitz. It is described by a power spectral density (PSD) $S(f)$ which is unknown. The objective is to estimate the regression vector $\beta$ that describe $\mu$ along with the PSD $S(f)$ that describes $\Sigma$ based on the observed entries of $y$. 

A way to solve this problem is to use methods like least-squares iterative adaptive approach \cite{Yardibi} which have been adapted to the missing data case by \citet{stoica2009missing} where they successively update the regression parameters and the covariance estimates. But this method requires to store and invert a $N_o \times N_o$ covariance matrix, where $N_o$ is the number of observed data. This involves the storage of $N_o^{2}$ values and $O(N_o^{3})$ operations for the inversion, which is challenging for large data sets. \citet{Baghi} proposed to construct an approximate general least-squares estimator, which avoids the direct storage and calculation of the covariance matrix. This is done by fitting an autoregressive (AR) model of order $p$ to the residuals, and using a Kalman filter in the de-correlation process, reducing the storage of the covariance parameters to $O(p)$ and the inversion to $O(pN)$ operations. While the regression has a quasi-minimal variance, the PSD estimate conditional to the AR model may be biased at low frequency. Therefore a less constrained estimator is desirable if a better estimate of the PSD is required.

An alternate approach is adopted by \citet{Berge2015} and \citet{Pires2016} where they estimate the missing values using a sparsity-prior with an algorithm commonly named \textit{inpainting} \cite{Donoho,Elad}, and then apply a standard least square regression on the reconstructed data. The precision and accuracy of the fit shows performance comparable to the KARMA method. While \textit{inpainting} is a way to fill the gaps prior to any regression method, it does not treat the Gaussian regression problem with missing data as a unified framework. However the idea of estimating the gaps to extract the parameters of interest from the data (and merely to have data series easier to handle) can be exploited differently as we will see next.

A general method to solve the Gaussian regression problem is the maximum likelihood estimation (MLE). When data are missing the direct maximization of the likelihood of the observed data is an optimization problem for which neither the solution nor the gradient of the cost function have an explicit form \cite{Szatrowski}. This can be circumvented by using iterative procedures where each step increases the likelihood and is convenient to calculate. 

The MLE problem can be solved by use of the expectation-maximization (EM) algorithm \cite{Dempster1977} which indirectly maximizes the likelihood of the observed data by computing the expectation of the likelihood of the complete data conditionally on the observed data. This involves the reconstruction of the missing data allowing the use of techniques adapted to regularly sampled data. However each iteration of the EM algorithm still requires $O(N_o^{3})$ operations in its original formulation. \citet{wang2005nonparametric} proposed an EM algorithm where the model is composed by harmonic functions and a given noise, whose covariance matrix is smoothed by partitioning the data vector $y$ in $L$ overlapping segments of size $M$. This procedure requires $O(LM^{3})$ operations at each iteration which can still be demanding if a high frequency resolution is required, and rely on the assumption that the sub-vectors are independent, which is not the case in general.

Storage and inversion of the observed data covariance matrix is a common issue in particular in the field of geostatistics where large spatial data sets may be considered. Several approaches are proposed, consisting in approximating the covariance matrix by a matrix close to it but easily invertible. Two strategies are usually adopted. 
The first one is covariance tapering (see e.g. \cite{Furrer}), where the autocovariance function is multiplied by a taper function which vanishes to zero after a certain number of points $q$. This introduces sparsity in the covariance matrix. The number of entries to be stored is reduced to $O(q^2)$ instead of $O(N_o^2)$ and the complexity of solving for linear equations involving $\Sigma$ is only linear in $N_o$, instead of being cubic. Another method is to perform a low-rank decomposition of the matrix, reducing the problem to $r^{2}$ parameters with $r \ll N_o$ (see e.g. \cite{Cressie}). The inversion can be performed with a complexity proportional to $O(r^{2}N_o)$.

Nevertheless these two methods suffer from drawbacks: covariance tapering captures the short range correlations only, whereas low-rank decomposition usually describes larger scales. In all cases, including the combination of both methods \cite{Sang}, this approach results in approximations which have proven to be inefficient for the spectral estimation problem relevant for MICROSCOPE.

More recently fast methods for solving linear systems based on circulant embedding of Toeplitz matrices have been proposed \cite{Miller,Fritz} enabling the exact resolution of MLE problems on incomplete data \cite{Stroud} with a reasonable computational complexity. But the authors assume that the autocovariance function - or equivalently the PSD - is well known or has a known form.

In this paper we tackle the problem of maximum likelihood estimation on time series with missing data and unknown, arbitrary power spectral density. The rationale relies on the assumption that the PSD is continuous and smooth, and that the complete data likelihood can be written in the circulant approximation (also known as Whittle's likelihood \cite{Whittle1951}). The procedure is based on a derivative of the EM algorithm called expectation-conditional-maximization (ECM) \cite{Meng&Rubin}, which is modified to introduce smoothing of the spectrum. This is done by a local formulation of the likelihood similarly to a formalism developed by \citet{Fan98automaticlocal} for spectral density estimation. This ECM-like algorithm is tailored to estimating $\beta$ and $S(f)$ and imputing the missing data, while the KARMA method \cite{Baghi} is used in the first place to find good starting values for the ECM algorithm. We then test our approach with numerical simulations.

In Sec.~\ref{sec:missing_problem} we first present the measurement model and we derive a general expression of the uncertainty versus the gap pattern and the noise PSD. In Sec.~\ref{sec:KARMA} we briefly review the KARMA methods which provides initial estimates for the regression parameters and the noise PSD in the presence of missing data. Then in Sec.~\ref{sec:data_reconstruction} we use these ``first guesses'' to estimate the missing data points. The main result of the paper is given in Sec.~\ref{sec:refinement} where we implement a method to refine the estimates of the regression parameters, the PSD and the reconstructed data based on a modified ECM algorithm, which is summarized in Sec.~\ref{subsec:final_ECM}. We finally test the method on simulated time series of the MICROSCOPE in-flight expected acceleration measurement in Sec.~\ref{sec:simulations}. Discussions and conclusions are given in Sec.~\ref{sec:conclusion}.

\section{\label{sec:missing_problem}Formulation of the missing data problem}
\subsection{\label{subsec:meas_eq}The measurement equation}
As many situations in experimental physics, the measurement equation can be written as a linear combination of measured or modeled signals. The aim of the data analysis process is to estimate the coefficients of this decomposition, and to characterize the residuals. 

In this framework, the measurement equation can be written in matrix form:
\begin{equation}
y = A \beta + z,
\label{eq:data_model}
\end{equation}
where $y = \begin{pmatrix} y_{0} & \ldots & y_{N-1} \end{pmatrix}^{T}$ is the $N \times 1$ vector containing the measured time series, regularly sampled at a frequency $f_{s} = 1 / \tau_{s}$ such that $y_n$ corresponds to time $t_n = n \tau_s$. 
$A$ is the $N \times K$ model matrix, i.e. the matrix containing all the measured or modeled signals that we want to fit to the data, and $\beta$ is the $K \times 1 $ vector of regression parameters to be estimated. The vector $z$ is a noise vector assumed to be a zero mean stationary Gaussian noise of unknown PSD that we denote by $S$.

Since some entries of $y$ may be missing, we define the window $w$ as a flagging vector such as $w_{n} = 1$ when the data are available at time $n\tau_s$ and $w_{n} = 0$ otherwise.

Let $y_{o}$ be the observed data vector that contains the valid data only. That is, if there are $N_{o}$ observed data at times $n_0,...,n_{N_{o}-1}$, the observation vector is equal to $y_{o} = \begin{pmatrix} y_{n_0} & \cdots & y_{n_{N_o-1}}\end{pmatrix}$, and we have $w_{n_i}=1$ for all  $0 \leq i < N_o$.

While this is not used for effective implementation in practice, it is also formally convenient to define the $N_o \times N$ indicator matrix $W_{o}$ of the observed data as:
\begin{eqnarray}
\forall \text{ } (i,j) \in \llbracket 0 ; N_o-1 \rrbracket \times \llbracket 0 ; N-1 \rrbracket, \\ \nonumber
W_o(i,j) = \left\{
    \begin{array}{ll}
        1 & \mbox{if } j = n_i, \nonumber \\ 
        0 & \mbox{otherwise.}
    \end{array}
\right.
\label{eq:obs_indicator_matrix}
\end{eqnarray}
The matrix $W_o$ is constructed such that $y_o = W_o y$. Likewise we construct the observed model matrix and the observed residuals vector as $A_o \equiv W_o A$ and $z_o \equiv W_o z$.

In a similar way we form the missing data vector $y_m$ and define the corresponding $N_m \times N$ indicator matrix such that $y_m = W_m y$. Also follow the definitions $A_m \equiv W_m A$ and $z_m \equiv W_m z$.

Finally, the effective regression problem is written by applying matrix $W_o$ on both sides of Eq.~(\ref{eq:data_model}):
\begin{equation}
y_o = A_o \beta + z_o,
\label{eq:observed_data_model}
\end{equation}
reflecting the fact that we perform a least-squares fit on the observed data only.
In the following we make explicit the ordinary least-squares solution of this system and its precision.

\subsection{Precision of the least-squares estimator as a function of PSD}
In a least-squares regression approach the uncertainty of the fit is directly linked to the noise PSD and the observation window $w$. We derive here a general expression of the estimation covariance as a function of the theoretical power spectral density.

The ordinary least-squares (OLS) solution of Eq.~(\ref{eq:observed_data_model}) for the regression vector is given by 
\begin{eqnarray}
	 \label{OLS_estimator}
   \hat{\beta} = (A_{o}^{\dag}A_{o})^{-1} \cdot A_{o}^{\dag} y_o,
\end{eqnarray}
where $\dag$ denotes the Hermitian conjugate. 

The covariance of the estimated parameter can then be written :
\begin{eqnarray}
   \mathrm{Cov}(\hat{\beta}) & = & Q_o^{-1}A_{o}^{\dag} \cdot W_o \Sigma W_o^{\dag} \cdot A_{o} Q_o^{-1},
	 \label{eq:covariance}
\end{eqnarray}
where we defined the matrix product $Q_o \equiv A_{o}^{\dag}A_{o}$ and the noise covariance matrix  $\Sigma \equiv \mathrm{E}\left[ zz^{\dag} \right]$. We can also define the covariance matrix of the observed noise as $\Sigma_{oo} \equiv W_o \Sigma W_o^{\dag}$. 

For a zero mean Gaussian stationary noise, in the discrete case, the random field is defined by its autocovariance function $R_{z}(\tau) \equiv \mathrm{E}\left[ z_{t} z_{t+\tau} \right]$. This even function is related to the power spectral density $S$ by the inverse Fourier transform:
\begin{equation}
R(\tau) = \int_{-\frac{f_{s}}{2}}^{\frac{f_{s}}{2}} S(f) e^{2 I \pi f \tau } df,
\label{eq:autocov_PSD}
\end{equation}
where $I=\sqrt{-1}$ is the complex number, and conversely the PSD is calculated from the autocovariance with the following relationship:
\begin{eqnarray}
S(f) = \frac{1}{f_{s}} \sum_{n = -\infty}^{\infty} R(n/f_{s}) e^{-  2 I \pi f n / f_{s}}.
\label{eq:autocorr_discrete}
\end{eqnarray}

As a result, the matrix $\Sigma$ can be written: 
\begin{eqnarray}
\Sigma(i,j) = R\left( (i - j) \tau_s \right) \text{ } \forall \text{ } (i,j) \in \llbracket 0 ; N-1 \rrbracket^{2}.
\label{eq:sigma_elements}
\end{eqnarray}
which defines a Toeplitz matrix. 
The integral in Eq.~(\ref{eq:autocov_PSD}) can be estimated at the sample times $n\tau_s$ by its equivalent Riemann sum:
\begin{equation}
\hat{R}(n\tau_s) = \frac{f_s}{P} \sum_{k=0}^{P-1}  S \left( f_k \right) e^{2I\pi \frac{nk}{P}},
\label{eq:autocov_PSD_estimate}
\end{equation}
for $P$ sufficiently large, where $f_k$ are the classical Fourier frequencies which read, if $P$ is even:
\begin{eqnarray}
f_k = \left\{
    \begin{array}{ll}
        kf_s/P & \mbox{if } 0 \leq k \leq P/2-1 \\ 
        (k-P)f_s/P & \mbox{if } P/2 \leq k \leq P-1.
    \end{array}
\right.
\label{eq:Fourier_freq}
\end{eqnarray}
The larger $P$, the more accurately the Riemann sum approximates the integral. To prevent any unphysical periodicity in $\hat{R}$, $P$ is generally chosen such that $P \geq 2N$. 

Eq. \ref{eq:autocov_PSD_estimate} allows us to write the covariance matrix in the following convenient way. 
We define $F_{P}(m,l) =  P^{-1/2} \exp{\left( -2 I \pi \frac{ml}{P} \right) }$ to be the $P \times P$ normalized discrete Fourier transform (DFT) matrix. We also define the $N \times P$ matrix $\Omega = \begin{pmatrix} I_{N} & 0_{P-N} \end{pmatrix}$ selecting the first $N$ entries of any vector by which $\Omega$ is multiplied. This is useful since we need $P$ values of the PSD to compute the $N$ values of the autocovariance. Then we can write the following set of equations:
\begin{subequations}
\label{eq:cov_matrix_as_function_of_PSD}
\begin{equation}
\Sigma = \Omega C \Omega^{\dag},\label{subeq:circulant_embedding}
\end{equation}
\begin{equation}
C = F_{P}^{\dag} \Lambda F_{P},\label{subeq:fourier_diagonalization}
\end{equation}
\end{subequations}
where $\Lambda$ is a $P \times P$ diagonal matrix defined by $\Lambda = f_s \times \mathrm{Diag}  \begin{pmatrix} S(f_0) & ... & S(f_{P-1}) \end{pmatrix}$.  
In the following we designate the matrix $\Lambda$ as ``the spectrum''. 
This formulation corresponds to the circulant embedding of Toeplitz matrices since $\Sigma$ is embedded in a larger circulant matrix $C$ (note that taking $P = N$ amounts to assuming that $\Sigma$ is circulant). This allows us to perform efficient calculations by use of fast Fourier transform (FFT) algorithms, as we show in Sec.~\ref{sec:large}. 

Finally Eqs. (\ref{eq:covariance}) and (\ref{eq:cov_matrix_as_function_of_PSD}) give the dependence of the OLS estimation error on the noise PSD and the observation window $w$. Next we review how the latter influences the error.

\subsection{\label{subsec:windows}Impact of the observation window}

If all data are available, all the entries of the window $w$ are equal to one. In that case $W_o$ is the identity matrix and the covariance of the observed data is a Toeplitz matrix.

If there are some gaps in the data, some entries of $w$ are zero. For an arbitrary gap pattern, the covariance of the observed data has entries $\Sigma_{oo}(i,j) = R\left((n_i-n_j)\tau_{s}\right)$, and has not a Toeplitz structure anymore. We showed in \cite{Baghi} that if the noise is colored (\textit{i.e.} the PSD is not constant) then the diagonal terms of the covariance of the ordinary least squares in Eq.~(\ref{eq:covariance}) can be significantly increased due to a convolution effect between the periodogram of the window and the noise PSD, leading to a leakage from high power regions to low power regions of the spectrum.

To prevent this increase of uncertainty, the estimator must be optimal with respect to the variance. We review our approach to build such an estimator in the next section.

\section{\label{sec:KARMA}Preliminary estimation of the regression parameters and the noise PSD: the KARMA method}

In order to avoid the increase of uncertainty of the regression in the presence of missing data, it is necessary to introduce a weighting based on the noise covariance in the estimator. This amounts to construct an approximation of the best linear unbiased estimator (BLUE) in terms of the variance.

In the ideal case where the noise PSD is known, the estimator with minimum variance is given by the general least-squares (GLS) estimator:
\begin{equation}
\hat{\beta} = \left( A^{\dag}_{o} \Sigma_{oo}^{-1} A_{o}\right)^{-1} A^{\dag}_{o} \Sigma_{oo}^{-1} y_{o},
\label{eq:GLS}
\end{equation}
where the observed noise covariance matrix admits a Cholesky decomposition $L$ such as $\Sigma_{oo} = LL^{\dag}$.

However, the exact noise covariance $\Sigma_{oo}$ is generally unknown and must be estimated. We implemented a method described in \cite{Baghi} which performs this estimation and use it to construct an estimator that approximates the GLS. 
The method (dubbed KARMA for Kalman AutoRegressive Model Analysis) iterates between 3 steps: first, the noise autocovariance is estimated by fitting an autoregressive (AR) model to the residuals $\hat{z}_{o} = y_{o} - A_{o}\hat{\beta}$. Secondly the orthogonalized vector $e_{o} = L^{-1}y_{o}$ is efficiently calculated with a Kalman filter. Thirdly $e_{o}$ is used to compute an approximate version of Eq.~(\ref{eq:GLS}).

At the end of the process, the KARMA method provides an efficient way to estimate the parameter vector $\beta$ with a precision close to the minimum variance bound (that would be obtained with exact GLS) without directly computing $\Sigma_{oo}^{-1}$. It also provides an estimate of the noise PSD $S$ calculated from the estimated autoregressive parameters. Note that all the process is done without estimating the missing data.

The KARMA outputs, that we denote $\hat{\beta}^{0}$ and $\hat{S}^{0}$ in the following, can be used to reconstruct the data in the missing intervals. We describe the reconstruction process in the next section.

\section{\label{sec:data_reconstruction}Data reconstruction using conditional expectations}
We saw in Sec.~\ref{sec:KARMA} that it is possible to estimate the regression parameters without filling the data gaps. However, performing missing data estimation (also called ``data imputation'') can be useful for two reasons. First, this is a good way to estimate the noise PSD more accurately. Indeed, while the accuracy of the estimated noise PSD $\hat{S}^{0}$ is sufficient to de-correlate the data and perform a precise regression for $\beta$, it may show a bias for certain shapes of $S$, especially at low frequencies.  
Secondly, equally spaced data sets are more easily considered for science purpose. The \textit{inpainting} method is a way to perform data imputation and allows for \textit{a posteriori} PSD estimation. However in this section we deal with a different approach called Gaussian interpolation which, for any set of estimate $\hat{\beta}$ and $\hat{S}$, allows us to estimate the missing data via their approximate conditional expectation.

\subsection{\label{subsec:description_reconstruction} Description of the data reconstruction process}

The indicator matrices introduced in Sec.~\ref{subsec:meas_eq} provide a convenient way to define the covariances of vectors $y_o$ and $y_m$ with themselves and with each other: 
\begin{eqnarray}
\Sigma_{oo} &\equiv& W_o \Sigma W_o^{\dag}, \nonumber\\ 
\Sigma_{mm} &\equiv& W_m \Sigma W_m^{\dag}, \nonumber\\ 
\Sigma_{mo} &\equiv& W_m \Sigma W_o^{\dag}
\label{eq:cov_matrix_with_indicator}.
\end{eqnarray}

By further assuming a Gaussian distribution of the noise, the optimal estimator of the missing data vector is its conditional expectation given the observed data vector:
\begin{eqnarray}
\mu_{m | o} &=& \mathrm{E} \left[ y_m | y_o , \beta, S \right] \nonumber\\ 
&=& \mu_{m} + \Sigma_{mo}\Sigma_{oo}^{-1}\left(y_{o}-\mu_{o}\right) 
\label{eq:mu_mis_given_obs_and_beta},
\end{eqnarray}
where the expectations of the missing and observed vectors are given by the regression model $\mu_{o} = A_{o} \beta$ and $\mu_{m} = A_{m} \beta$. In our application the first term in Eq.~(\ref{eq:mu_mis_given_obs_and_beta}) represents the deterministic part of the reconstruction, whereas the second term accounts for the stochastic noise statistics.

The mean squared prediction error (MSE) is equal to the conditional covariance $\Sigma_{m|o}$ of $y_m$ given $y_o$. If $\epsilon_m = y_m - \mu_{m | o}$ is the residual of the missing data estimation then the MSE read
\begin{eqnarray}
\mathrm{E} \left[ \epsilon_m \epsilon_m^{\dag} \right] &=& \Sigma_{m|o} \nonumber\\ 
&=& \Sigma_{mm} - \Sigma_{mo}\Sigma_{oo}^{-1} \Sigma_{mo}^{\dag}
\label{eq:MSE_mis_given_obs_and_beta}.
\end{eqnarray}

In our study however, both $\Sigma$ and $\beta$ are assumed to be unknown. They must therefore be determined beforehand from the observed data. For instance, one can use the estimates provided by the KARMA method. We can then replace the expectations in Eq.~(\ref{eq:mu_mis_given_obs_and_beta}) by the estimated ones, that is $\hat{\mu}_{o} = A_{o} \hat{\beta}$ and $\hat{\mu}_{m} = A_{m} \hat{\beta}$. 

As for the covariances involved in Eq.~(\ref{eq:mu_mis_given_obs_and_beta}), they are replaced by their estimate $\hat{\Sigma}$, derived from the estimate of the PSD $\hat{S} = \hat{S}^{0}$ and by using Eqs (\ref{eq:cov_matrix_as_function_of_PSD}) and (\ref{eq:cov_matrix_with_indicator}).

The uncertainty introduced by the estimation of $\beta$ leads to an additional error term in the reconstruction \cite{Harville}:
\begin{eqnarray}
\mathrm{E} \left[ \epsilon_m \epsilon_m^{\dag} \right]  = \Sigma_{m|o} + K_m \mathrm{Cov}\big(\hat{\beta}\big) K_m^{\dag}
\label{eq:MS_mis_given_obs},
\end{eqnarray}
with $K_m \equiv A_m -  \Sigma_{mo} \Sigma_{oo}^{-1} A_o$.
The uncertainty of the spectrum estimate $\hat{S}$ also affects the reconstruction error. However the full derivation of the corresponding error term is beyond the scope of this study. See for instance \cite{Harville2} for a discussion on this aspect.

\subsection{\label{sec:large}The preconditioned conjugate gradient method}
As pointed out in section \ref{subsec:windows}, the covariance matrix of the observed data $\Sigma_{oo}$ looses its useful Toeplitz properties in the presence of missing data, and cannot be inverted exactly and efficiently (see for instance \cite{Ammar1996,Martinsson2005} for superfast Toeplitz inversion algorithms). For time series with $10^{6}$ data points, it is not feasible to store and invert the $N_o \times N_o$ covariance matrix. 

In the analysis of stationary time series that are originally regularly sampled, the covariance matrix of the observed data is only defined by two vectors of size $N$: the window $w$ and the PSD $S$. The product of this matrix by any vector can be calculated using FFT algorithms and element-wise vector multiplications, taking advantages of the covariance form described by Eqs. (\ref{eq:cov_matrix_as_function_of_PSD}) and (\ref{eq:cov_matrix_with_indicator}). As a result, the linear system $\Sigma_{oo} x = y$ involved in the estimation of the missing data in Eq.~(\ref{eq:mu_mis_given_obs_and_beta}) can be efficiently solved by iterative algorithms decreasing the residuals $r_l = y - \Sigma_{oo} x_l$ at each iteration $l$ leading to an approximate (and sometimes exact) solution. For example \citet{Fritz} suggest to use the conjugate gradient algorithm \cite{Stiefel1952}. However this process may be slow, and the use of a preconditioner matrix $M$ is required to reduce the condition number of the linear system, which amounts to solving $M^{-1} \Sigma_{oo} x = M^{-1} y$. 

While \citet{Fritz} use the regularized circulant preconditioner introduced by \citet{Nowak05}, we choose a tapered covariance as the preconditioner matrix, in order for $M$ to have the same non-Toeplitz structure as $\Sigma_{oo}$. To contruct $M$, the correlation is ignored after a certain lag $\tau_0$ by mutiplying the original autocovariance by a taper function $K$ which smoothly goes down to zero at lag $\tau_0$:
\begin{equation}
R_{\rm{taper}}(\tau) = R(\tau) K(\tau,\tau_0).
\label{eq:tapered_covariance}
\end{equation}
The resulting covariance matrix is sparse which allows us to store it and solve the corresponding linear system with an acceptable number of operations.

The preconditioning step adds two operations which are linear in $N_o$ at each iteration of the conjugate gradient algorithm. The resulting reconstruction process has a complexity in $O(N_{\rm{it}}N \log N)$, where $N_{\rm{it}}$ is the number of iteration needed by the conjugate gradient to reach convergence.

\section{\label{sec:refinement}Refinement of regression parameters and noise spectrum estimation through a modified ECM algorithm}

In the previous section we showed how to infer the missing data from given estimates of the regression vector $\hat{\beta}$ and of the PSD $\hat{S}$. But it is possible to re-estimate these quantities from the reconstructed data, and to iterate between imputation and estimation steps. The objective of such an iterative process is two-fold: on the one hand to improve the estimation of the noise PSD, and on the other hand to obtain a consistent reconstruction of the missing data. These iterations are implemented via a modified ECM algorithm which starts from the initials guesses $\hat{\beta}^{0}$ and $\hat{S}^{0}$ provided by the KARMA method and perform several reconstruction/estimation steps to end up with a converged set of estimates $\hat{y}_m$, $\hat{\beta}$ and $\hat{S}$.

\subsection{\label{subsec:log-likelihood}Approximate likelihood for complete data}
We first examine the case where all data are available. Let $\theta = \left( \beta,S\right)$ be the set of parameters to estimate (here ``$S$'' refers to the set of parameters necessary to describe the PSD, which is given by the $\Lambda$ matrix). For the regression model in Eq.(\ref{eq:data_model}) with $W = I$ and with Gaussian stationary residuals, the log-probability density $l_y(\theta)$ of the full data (\textit{i.e}, the logarithm of the probability to observe $y$ given $\theta$) can be written as:
\begin{eqnarray}
l_y(\theta) &=& \log p(y|\theta) \nonumber \\
&=& - \frac{1}{2} \Big\{ \log(2\pi) + \log |\Sigma| \nonumber \\ 
& & + \left( y - A\beta \right)^{\dag} \Sigma^{-1} \left( y - A\beta \right) \Big\}.
\label{eq:likelihood}
\end{eqnarray}

It can be shown that for large $N$ the decomposition of the covariance matrix in Eq.~(\ref{eq:cov_matrix_as_function_of_PSD}) can be approximated by:
\begin{equation}
\Sigma \approx F_{N}^{\dag} \Lambda F_{N},
\label{eq:sigma_approx}
\end{equation}
which means that the covariance matrix is approximately diagonalisable in Fourier space (see e.g. \cite{Whittle1951,Whittle1957}). Under this hypothesis the log-likelihood can be re-written as
\begin{equation}
l_y(\theta) \approx - \frac{1}{2} \sum_{k=0}^{N-1} \left\{ \log \Lambda_k  + \frac{\left| \tilde{y} - \tilde{A}\beta \right|^{2}}{\Lambda_k}  \right\},
\label{eq:log-likelihood_Fourier}
\end{equation} 
where we dropped the constant terms and we respectively defined the normalized DFT of the data and of the model matrix by $\tilde{y} = F_N y$ and $\tilde{A} = F_N A$. We note that the log-likelihood involves the periodogram $I_{z}$ of the residuals $z = y - A\beta$, where we define the periodogram of any vector $x$ as the squared modulus of its normalized DFT:
\begin{equation}
I_{x} \equiv \left| \tilde{x} \right|^{2}
\label{eq:def_periodogram}
\end{equation}

\subsection{\label{subsec:ECM}ECM algorithm for missing data}
In the case where data are missing, the log-likelihood no longer has the simple form given by Eq.~(\ref{eq:log-likelihood_Fourier}). Then it is not easy to estimate the power spectral density of the residual process through direct maximization of the log-likelihood function. Indeed this would imply to solve a non-linear optimization problem with $P+K$ parameters, where $P$ is the number of parameters necessary to describe the PSD. However when data are regularly sampled, it is possible to use fast spectral estimation techniques. The expectation maximization (EM) algorithm developed by \citet{Dempster1977} is a way to address this issue. 

The EM algorithm is an iterative procedure which maximizes the likelihood of the observed data $y_o$ with respect to the model parameters. The idea of each iteration $i$ is to first estimate the expectation of the complete data log-likelihood $l_{y}$ conditionally on the observed data and on the current estimate of the model parameter $\theta^{i}$. Then the second step updates the model parameter $\theta$ by maximizing the likelihood with respect to them. These two steps (labeled E and M) are repeated until convergence of the parameters. 

The E-step is the computation of the conditional expectation of the log-likelihood: $l_{y|o}(\theta) = \mathrm{E}\left[ l_{y}(\theta) |y_o,\theta^{i} \right]$. It involves two sub-steps. The first one, called E1, is the estimation of the missing data given by Eq.~(\ref{eq:mu_mis_given_obs_and_beta}) from which we obtain a reconstructed data vector $y^{i} = \mathrm{E}\left[y|y_o,\theta^{i} \right]$. The second sub-step, called E2, is the estimation of the conditional second order moment $\mathrm{E}\left[yy^{\dag}|y_o,\theta^{i} \right]$, and more practically the conditional periodogram. This sub-step is detailed in Sec.~\ref{subsec:periodogram_expectation}. 

The M step is the maximization of the conditional expectation $l_{y|o}(\theta)$. It can be computationally expansive to directly maximize $l_{y|o}(\theta)$, therefore we use a subclass of EM algorithm called expectation-conditional-maximization (ECM). This algorithm shares the same properties as the EM procedure (see e.g. \cite{Meng&Rubin} and \cite{LittleRubin}, pp 179-181). In this version, the M step does not maximize $\mathrm{E}\left[ l_{y}(\theta) |y_o,\theta^{i} \right]$ but simply increases it and is usually divided in several sub-steps. In the Gaussian regression problem described by Eq. \ref{eq:log-likelihood_Fourier} two sub-steps are necessary and correspond to conditional and successive updates of the vector $\beta$ and the spectrum $\Lambda$. 
These two steps are labeled CM1 and CM2 hereafter. At iteration $i+1$ they read:
\begin{itemize}
	\item CM1 step:
\begin{equation}
\beta^{i+1} = \left( \tilde{A}^{\dag} \left(\Lambda^{i}\right)^{-1} \tilde{A} \right)^{-1} \tilde{A}^{\dag} \left(\Lambda^{i}\right)^{-1} \tilde{y}^{i}
\label{eq:CM1}
\end{equation}
	\item CM2 step:
\begin{equation}
\Lambda_k^{i+1} = \left| \tilde{y}^{i}_k - \left( \tilde{A}\beta^{i+1} \right)_{k} \right|^{2}
\label{eq:CM2}
\end{equation}
\end{itemize}
were $\tilde{y}^{i}$ is the DFT of the reconstructed data vector $y^{i}$.

Note that Eqs. (\ref{eq:CM1}) and (\ref{eq:CM2}) are valid under the circulant hypothesis (\ref{eq:sigma_approx}). If the number of points $N$ is not large enough for this assumption to be acceptable, the residuals $z$ can be considered to be a sub-sample of a larger times series $z^P$ of size $P$ having a circulant covariance given by Eq.~(\ref{subeq:fourier_diagonalization}) as suggested in \cite{Miller,Stroud}. In that case the data $y_m^{P}$ considered as ``missing'' include the vector $y_m$ defined in Sec.~\ref{sec:missing_problem} plus a vector of size $P-N$. The indicator matrix for the observed data is then completed with zeros to form the effective indicator matrix $W_o^{P} = \begin{pmatrix} W_o & 0_{N_o \times P-N } \end{pmatrix}$. However in the following we assume that the circulant hypothesis is valid and that the log-likelihood of complete data has the form given by Eq.~(\ref{eq:log-likelihood_Fourier}).

Besides, the particularity of the uni-variate problem is that we do not have several realizations of $y_o$ which we can average as it is usually assumed in standard utilization of the EM algorithm. The PSD estimate $\Lambda$ as written in Eq.~(\ref{eq:CM2}) will therefore have a large variance. In order to reduce the variance, the CM2 step must be modified by introducing smoothing. We construct such a modified estimator in the next section.

\subsection{\label{subsec:PSD_estimation}Modification of the CM2 step: estimation of the noise power spectral density by spectrum smoothing}
We now specify the PSD estimator needed to introduce smoothing in the CM2 step of the ECM algorithm. 
A classical approach is to use the Welch's method \cite{Welch} or similar periodogram smoothing techniques reviewed by \citet{priestley1982spectral}. However, given that we aim to use the PSD estimate to reconstruct the data, we would like it to fulfill the following requirements: 1/ it must have a small variance and be sufficiently smooth; 2/ it must have a low bias at low frequencies (since we want to accurately estimate the error in this region) 3/ be quickly computable (for $N \sim 10^{6}$).

An ideal candidate is the local linear model, whose adaptation to spectral estimation is described in \cite{FanYao} and \cite{Fan98automaticlocal}. The driving idea is to assume that locally, the log-spectrum can be described by a linear function. Let $f_j$ be some frequency in the interval $\left[ 0 , f_s/2 \right]$ at which we want to estimate the PSD. This amounts to saying that if $f_k$ is a frequency in the neighborhood of $f_j$, then we have:
\begin{equation}
\log \Lambda_k \approx a_j + b_j (f_k-f_j),
\label{eq:linear_approx_loglikelihood}
\end{equation}
where $a_j$ and $b_j$ are coefficients to be estimated for each frequency $f_j$. The extent of the neighborhood is determined by a kernel $\mathrm{K}$ and a specific bandwidth $h$, which can depend on $f_j$.

The advantage of using such an estimator is two-fold.
First, this estimator is nearly unbiased for estimating steep trends, while standard kernel smoother usually show a large bias when the spectrum is steep near the boundary. Second, for applications in physics it makes sense to apply this kind of smoothing to the log-periodogram because the shape of the spectrum is likely to be locally a power law function. 
That being said, the local linear estimator can be used for very general shapes.

In the framework of maximum likelihood estimation this can be expressed as a local formulation of the likelihood which we denote ${l_y}^j$. For a frequency $f_j$ of interest we re-write Eq.~(\ref{eq:log-likelihood_Fourier}) by using the linear model of Eq.~(\ref{eq:linear_approx_loglikelihood}) and by applying a weight for each frequency:
\begin{eqnarray}
{l_y}^j(a,b) &=& - \frac{1}{2} \sum_{k=1}^{n} \Big\{ a + b(f_k-f_j) \nonumber \\  
&& + I_{z}(k) e^{-a - b (f_k-f_j)} \Big\} \mathcal{K}_{h_j}(f_k - f_j),
\label{eq:local_likelihood}
\end{eqnarray} 
where we have restricted the summation to the positive frequencies, $k$ running from $0$ to $n = \lfloor{ (N-1)/2 \rfloor}$. The smoothing in the frequency domain is driven by the weight function $\mathcal{K}_h(f) = \frac{1}{h} \mathcal{K}(f/h)$ where $\mathcal{K}(x)$ is a given kernel vanishing to zero for $x \rightarrow \infty$. 

The estimation of the covariance parameter $\Lambda_j$ at frequency $f_j$ is given by the local intersect $a_j$ that maximizes Eq.~(\ref{eq:local_likelihood}). 

The solution of this maximization is exactly the same as the local maximum likelihood estimator proposed by \citet{Fan98automaticlocal}, and can be solved with the Newton-Raphson algorithm \cite{Newton} in a single iteration.

We could maximize Eq.~(\ref{eq:local_likelihood}) for all Fourier frequencies, \textit{i.e.} taking the $f_j$'s equal to the $f_k$'s in Eq.~(\ref{eq:Fourier_freq}). However, to reduce the computational cost of the PSD estimation the maximization of Eq.~(\ref{eq:local_likelihood}) is performed only for frequencies $f_j$ on a down-sampled grid of size $J$, with $J \ll N$. Then the values of the PSD necessary to compute the spectrum matrix $\Lambda$ are obtained by linear interpolation. 
For the interpolation to be precise within all the frequency decades of the measurement bandwidth, the initial frequency grid is taken to be logarithmic, as well as the smoothing parameter $h_j$, in a way similar to the LPSD method proposed by \citet{Trobs}. Thus the $f_j$'s are chosen such that $f_j = (f_s/P) \left( P/2 \right)^{j/(J-1)}$.

Furthermore, the number of operations to compute the maximizer of the local log-likelihood is reduced by choosing $\mathcal{K}(x)=0$ for $x>1$. In this paper we use the Epanechnikov kernel \cite{Epa69}, which is optimal in the sense of the mean integrated squared error (MISE):
\begin{eqnarray}
\mathcal{K}(x) = \left\{
    \begin{array}{ll}
        \frac{3}{4} \left( 1 - x^2 \right)  & \mbox{if } |x| \leq 1, \nonumber \\ 
        0 & \mbox{otherwise.}
    \end{array}
\right.
\label{eq:epanechnikov}
\end{eqnarray}

Up to now we have written the local log-likelihood in the case of complete data. However in the ECM algorithm the update of the spectrum estimate is performed conditionally on the observed data. This is easily done by taking the conditional expectation of Eq.~(\ref{eq:local_likelihood}) with respect to $y_o$, and then maximizing it with respect to $(a,b)$. It follows that the maximization of the likelihood at iteration $i$ involves the conditional expectation of the periodogram $I_{z}^{i} = \mathrm{E}\left[I_{z}|y_o,\theta^{i} \right]$. We detail its computation below.

\subsection{\label{subsec:periodogram_expectation}Conditional expectation of the periodogram}
 
Given the definition of the periodogram in Eq.~(\ref{eq:def_periodogram}), its conditional expectation involves cross-products of the data vector $y$, hence requiring the calculation of the conditional covariance of $y$ given $y_o$. 
The formal derivation of the conditional expectation of the periodogram is performed in Appendix \ref{appendix:cond_per}, and reads
\begin{equation}
\mathrm{E}\left[ I_z(k) |y_o, \theta \right] =  I_{\hat{z}}(k) + \sigma_k^{2}.
\label{eq:cond_periodogram}
\end{equation}

The first term of this equation is the periodogram of the reconstructed residual vector $\hat{z}$ whose entries are equal to $z_n$ when the data are observed at time $t_n$, and equal to the conditional expectation of $z_n$ given $y_o$ and $\theta^{i}$ when the data are missing at $t_n$: $\hat{z} = W_o^{\dag} z_o + W_m^{\dag} \mu_{m|o}$, where $\mu_{m|o}$ is given by Eq.~(\ref{eq:mu_mis_given_obs_and_beta}).

The second term accounts for the conditional second order moments. The quantities $\sigma_k^{2}$ are given by the entries of the vector: $\sigma^{2} \equiv \mathrm{diag} \left( F_{N} W_{m}^{\dag} \Sigma_{m|o} W_{m} F_{N}^{\dag} \right)$, where $\Sigma_{m|o}$ is the conditional covariance in Eq.~(\ref{eq:MSE_mis_given_obs_and_beta}). For large numbers of missing data points, the direct calculation of this term can be computationally demanding, since it involves the matrix product $\Sigma_{oo}^{-1}\Sigma_{om}$, whose complexity is of order $O\left(N_{\rm{it}} N_m N_o \log N_o\right)$ using the conjugate gradient algorithm of Sec.~\ref{sec:large}.

To avoid these computational issues $\sigma^{2}$ can be approximated with Monte Carlo simulations of the vector $\hat{z}$ as described in Appendix \ref{appendix:cond_generation}. 
The idea is to generate several realizations of the reconstructed vector $\hat{z}$, to calculate their periodograms, and to compute their sample average.
The complexity can then be reduced to $O\left(N_{\rm{it}} N_d N_o \log N_o\right)$ where $N_d$ is the number of Monte-Carlo draws. This alternative is obviously of interest when $N_d < N_m$. 
We show below (Sec.~\ref{subsec:PSD_results}) that for our application a rough approximation of $\sigma^{2}$ (\textit{i.e.} low $N_d$) is sufficient.

\subsection{\label{subsec:error_assessment}Precision assessment} 
For a given observed datum $y_o$, once the PSD is estimated by the modified ECM (M-ECM) algorithm, the uncertainty of the regression parameters can be approximately assessed. This is very useful since in practice only a few realizations of $y_o$ are available. The error is evaluated by estimating the covariance of the GLS estimator defined in Eq.~(\ref{eq:GLS}), \textit{i.e.}:
\begin{equation}
\mathrm{Cov}\big(\hat{\beta}\big) = \left( A_o^{\dag} \Sigma_{oo}^{-1} A_o\right)^{-1},
\label{eq:GLS_covariance}
\end{equation}
where $\Sigma_{oo}$ is replaced by its M-ECM estimate, giving an approximate assessment of the precision of the regression. 
The calculation of Eq.~(\ref{eq:GLS_covariance}) requires to solve $K$ linear systems involving the matrix $\Sigma_{oo}$. Similarly to the computation of the conditional missing data estimate in Sec.~\ref{sec:large}, their solutions are obtained by using an iterative method involving matrix-vector products only. However for this calculation the PCG algorithm shows an irregular convergence behavior. We use the Biconjugate gradient stabilized method (BiCGSTAB) \cite{Lanczos52solutionof} instead. While adapted to the more general case of non-Hermitian matrices (unlike the PCG) this variant has better stability properties. To accelerate convergence a preconditioned version of the BiCGSTAB method is used \cite{Hogben}, with the same sparse preconditioner as in Sec.~\ref{sec:large}. The iterations are stopped when the residuals $r$ have decreased by some specified amount. 

The estimate of the standard deviation of $\hat{\beta}$ is obtained by taking the square root of the diagonal of the estimated GLS covariance in Eq.~(\ref{eq:GLS_covariance}). In the following we call this the ``M-ECM error estimate'' and denote it $\hat{\sigma}_{\rm{ECM}}$.

\subsection{\label{subsec:final_ECM}Final formulation of a modified ECM algorithm}
This section summarizes the main steps of the likelihood maximization procedure. An iteration $i$ of the developed algorithm can be summarized as follows:

\begin{itemize}
\item \textit{Initialization:} calculation of the first guesses $\hat{\beta}^{0}$, $\hat{S}^{0}$ with the KARMA method described in Sec.~\ref{sec:KARMA};
\item \textit{E step: calculation of the terms involved in the conditional expectation of the likelihood}
\begin{enumerate}[label=E\arabic*,start=1]
	\item Calculation of the conditional data vector given the observed data $y^{i} = \mathrm{E}\left[y|y_o,\theta^{i} \right]$ with Eq.~(\ref{eq:mu_mis_given_obs_and_beta});
	\item Calculation of conditional periodogram given the observed data $I_{z}^{i} = \mathrm{E}\left[I_{z}|y_o,\theta^{i} \right]$ following the method described in Sec.~\ref{subsec:periodogram_expectation} which gives an approximation of Eq.~(\ref{eq:cond_periodogram});
\end{enumerate}
\item \textit{CM step: conditional maximization of the likelihood}
\begin{enumerate}[label=CM\arabic*,start=1]
	\item Calculation of the new estimate $\beta^{i+1}$ with Eq.~(\ref{eq:CM1}).  
	\item Calculation of the new estimate of the spectrum $\Lambda^{i+1}$. To do this, we first compute the reconstructed residuals $z^{i} = y^{i} - A \beta^{i+1}$. Then, after replacing the periodogram $I_{z}$ by its conditional estimate $I_{z}^{i}$ in Eq.~(\ref{eq:local_likelihood}), ${l_y}^j$ is maximized with respect to $(a,b)$ for each frequency $f_j$ to obtain $(a_j,b_j)$, where the spectrum estimate is $\Lambda^{i+1} = a_j$. The values of the spectrum at all Fourier frequencies $f_k$ are deduced by linear interpolation from the values at $f_j$.
\end{enumerate}
\end{itemize}

It must be noted that this algorithm is not exactly an ECM because we introduced a local maximization at the CM2 step (Sec.~\ref{subsec:PSD_estimation}) so that the form of the likelihood slightly changes with respect to the CM1 step. That is why we designate it as M-ECM.

We apply the M-ECM algorithm to data simulated in the framework of the MICROSCOPE space mission in Sec.~\ref{sec:simulations}.

\section{\label{sec:simulations}Numerical tests}

After briefly describing the MICROSCOPE space experiment and the associated measurement model, we apply the M-ECM method to mock in-flight data. We successively analyze the result of the missing data reconstruction, the PSD estimation, the regression parameters estimation, and the error assessment.

\subsection{The MICROSCOPE space mission}
In the current efforts to build a harmonized theory of both cosmic and quantum scales, some models such as string theory postulate the existence of new interactions leading to a violation of the weak equivalence principle (WEP). In particular, some works predict a violation of the universality of free-fall at the $10^{-13}$ level \cite{Damour}, in conflict with general relativity. To provide experimental data to these theoretical investigations, the MICROSCOPE space mission is designed to test the WEP with a precision of $10^{-15}$ \cite{Touboul}, two orders of magnitude better than current on-ground experiments \cite{wagner2012torsion,Will}. The satellite follows a quasi-circular and sun-synchronous orbit around Earth and carries two differential electrostatic accelerometers. Each of them probes the free-fall of two test-masses (TMs). In the first accelerometer, called EP sensor unit, the composition of the two TMs is different: one is made of Platinum Rhodium alloy (PtRh), whereas the other is made of Titanium alloy (TA6V). In the second accelerometer, called REF sensor unit, the two TMs are both made of PtRh. This instrument thus serves as an experimental reference. To finely monitor the free-fall, the TMs are servo-controlled by a set of electrodes so that they stay relatively motionless at the center of the accelerometers cages. The signal of interest in the experiment is the difference between the accelerations applied to the two TMs which is deduced from the electrostatic force needed to maintain them at the center of the cages. Because of the instrument imperfections, the differential acceleration is coupled to the mean acceleration of the two TMs. This effect is nullified by a drag-free system implemented in the satellite. 

The main objective is to estimate the E\"{o}tv\"{o}s parameter defined for two TMs labeled 1 and 2 as the ratio $\delta \equiv 2\left(\frac{m_{g,1}}{m_{i,1}} - \frac{m_{g,2}}{m_{i,2}}\right)/\left(\frac{m_{g,1}}{m_{i,1}} + \frac{m_{g,2}}{m_{i,2}}\right)$ where $m_i$ and $m_g$ respectively refer to the inertial and the gravitational mass. A violation of the WEP would be visible in the difference between the accelerations of the TMs measured by the EP sensor unit and would be approximately proportional to the E\"{o}tv\"{o}s parameter.

During the WEP test sessions (\textit{i.e.} the time intervals of the mission when the test is performed) the satellite attitude is finely controlled in order to either point to a constant direction (inertial session) or to slowly rotate the instrument about the axis normal to the orbital plane (spin session). This pointing constrains the possible WEP violation signal to have a specific signature with a frequency $f_{\rm{EP}}$ equal to the sum of the orbital frequency $f_{\rm{orb}}$ and the spin frequency of the satellite $f_{\rm{spin}}$. In order to get a sufficient signal-to-noise ratio the measurement is integrated during several orbits. 

\subsection{\label{subsec:data_sample}MICROSCOPE data}
\subsubsection{\label{subsec:data_sample}Model}

We apply the M-ECM algorithm to representative simulated measurement of the MICROSCOPE accelerometers. The time series are generated with a mission simulator and are the same as in \cite{Baghi}. They correspond to a spin session sampled at $f_s = 4 \text{ }\rm{Hz}$ and lasting 20 orbits, which is the time necessary to ensure a precision of at least $10^{-15}$ on the E\"{o}tv\"{o}s parameter for a complete measurement. The duration of one session is about 33 hours, resulting in $N = 4.7 \times 10^{5}$ data points for a complete series. In the formalism of Eq. \ref{eq:observed_data_model}, the observed data vector is defined by the differential acceleration $y_o = W_o \gamma_d$ at the observed times, where $\gamma_d$ is the half-difference of the accelerations of the two TMs. For this study, the signal is simplified and contains 3 components only: a term proportional to the gravitational acceleration projection $g_x$ on the sensitive axis of the instrument ($x$-axis) resulting from a possible WEP violation, and two terms proportional to the gravitational gradients $T_{xx}$ and $T_{xz}$ appearing because the TMs are not perfectly centered with respect to one another. Hence there are 3 regression parameters to estimate: the E\"{o}tv\"{o}s parameter $\delta$ and the off-centerings $\Delta_x$ and $\Delta_z$. The model matrix and the regression parameter read:
\begin{eqnarray}
A &=& \frac{1}{2} \begin{pmatrix} g & T_{xx} & T_{xz} \end{pmatrix}; \nonumber \\
\beta &=& \begin{pmatrix} \delta & \Delta_x & \Delta_z \end{pmatrix}^{\dag},
\label{eq:model_for_simus}
\end{eqnarray} 
where the $1/2$ factor is simply due to the definition of $\gamma_d$. In the following a WEP violation is simulated with $\delta = 3 \times 10^{-15}$ and the off-centerings are set equal to $\Delta_x = \Delta_z = 20 \text{ }\rm{\mu m}$.
The main harmonic of the gravitational acceleration signal $g_x$ is located at the WEP frequency $f_{\rm{EP}} = 9.35 \times 10^{-4}$ Hz, whereas the main harmonic of the gravitational gradient perturbations is found at $2f_{\rm{EP}} = 1.87 \times 10^{-3}$ Hz.

\subsubsection{\label{subsec:gaps}Gap patterns}
As mentioned in Sec.~\ref{sec:intro}, missing data in the MICROSCOPE space mission might come from tank or MLI crackles inducing saturations of the accelerometer measurement. At the time of writing MICROSCOPE is not launched yet, and the exact probability of occurrence of these events is currently unknown, but worst case figures have been estimated by on-ground tests and are used in the simulations presented here.

In this study we deal with three kinds of observation windows: complete data, randomly distributed gaps (``random gaps'' for short) and periodic gaps. Only OLS and KARMA estimations are run when data are complete, serving as a comparison for the results obtained in the presence of gaps.

The random window simulates random and unpredictable saturations in flight, such as tank crackles perturbations. In this case we assume that the times at which the gaps begin is a random variable following a discrete uniform distribution on the interval $\llbracket 0 ; N-1 \rrbracket$, with 260 gaps per orbit and a duration of 0.5 seconds for each gap. 

The periodic window simulates the data unavailability that could occur at a special frequency (due to periodic temperature changes for example). In this case there exists a period $T_g$ such that $w_{n+T_{g}f_{s}} = w_{n}$. The period of the interruptions chosen in the simulation is the orbital period $T_{\rm{orb}} = 1/f_{\rm{orb}}$ which is a likely periodicity for an experiment on-board an orbiting satellite. Both windows are sized such that they represent a 2\% fraction of data losses, or about $N_m = 10^{4}$ points. As a result, the random gaps are shorter and more frequent than the periodic gaps across the time series.

\subsubsection{\label{subsec:residuals}Noise model}
We draw 400 realizations of the noise vector $z$ generated from a PSD $S(f)$ according to the method described by \citet{timmer1995generating} (this number of draws is chosen to have an error less than $10^{-16}$ on the sample standard deviation of $\hat{\delta}$ with a 99\% confidence). The PSD corresponds to a physical model of the accelerometer noise which can be approximated by a power law and a transfer function $H$:
\begin{equation}
S(f) = \left( \alpha_{0} + \alpha_{-1} f^{-1} + \alpha_{4} f^{4} \right) \left|H(f)\right|^{2}, 
\label{eq:PSD_model}
\end{equation}
where $H$ accounts for the attenuation in high frequencies due to the control loop of the accelerometers and the anti-aliasing filters.

\subsection{\label{subsec:algorithm_parametrization}Parametrization of the modified ECM algorithm}

In this section we give some details about how the M-ECM algorithm is parametrized.

For each draw of $z$ the observed vector $y_o$ is constructed according to Eq.~(\ref{eq:observed_data_model}). After obtaining the KARMA estimates  $\hat{\beta}^{0}$ and $\hat{S}^{0}$, the M-ECM algorithm is run to obtain the final estimates $\hat{\beta}$, $\hat{S}$, and $\hat{y}_m$. This scheme is repeated for the 400 draws.

The iterations of the M-ECM algorithm is stopped when the difference between the current and the previous estimation of the E\"{o}tv\"{o}s parameter $\hat{\delta}_{i} - \hat{\delta}_{i-1}$ is less than $10^{-17}$.

At each iteration $i$ of the ECM, the preconditioned conjugate gradient (PCG) algorithm described in Sec.~\ref{sec:large} is run to calculate $x = \Sigma_{oo}^{-1} z_o$. This process itself involves $N_{\rm{it}}$ iterations, with typically $N_{\rm{it}} \sim 100$. The taper length $q = \tau_0 f_s$ for the preconditioner is selected to be equal to the order of the AR process used in the KARMA estimation. The latter is found by minimizing the Akaike information criterion determined from the data, namely $p = 60$ in the case of the simulated noise (see \cite{Baghi} for more details). The choice $q=p$ allows us to represent the noise correlation to a range sufficient to approximate the GLS estimator in the KARMA method. The iterations of the PCG algorithm are stopped when the norm of the solution residuals $r = z_o - \Sigma_{oo} x$ reaches the threshold $\epsilon$, chosen to be the standard deviation of the residuals $z_o$ of the observed data model.

At each iteration, the conditional periodogam given by Eq.~(\ref{eq:cond_periodogram}) is approximated by $N_{d} = 5$ Monte-Carlo draws. $N_d$ is taken to be sufficiently low to reduce the computational cost. We will see next that this choice has not a major incidence on the result. Nevertheless when processing real experimental data for which few realizations are usually available, it is obviously safer to use a higher number of conditional draws.

The algorithm is implemented in python language and the DFTs are computed using the python wrapper around the efficient FFTW library \cite{FFTW}. Each iteration of the M-ECM procedure takes almost 3 minutes for $N \sim 5 \times 10^{5}$ on a typical 2 GHz computer. For the data under study the convergence is obtained within less than 10 iterations.

\subsection{\label{reconstruction_results}Results of a single reconstruction}
We present here the result of a reconstruction obtained on a single simulation.

\begin{figure}%
\flushleft
\includegraphics[width=0.9\columnwidth]{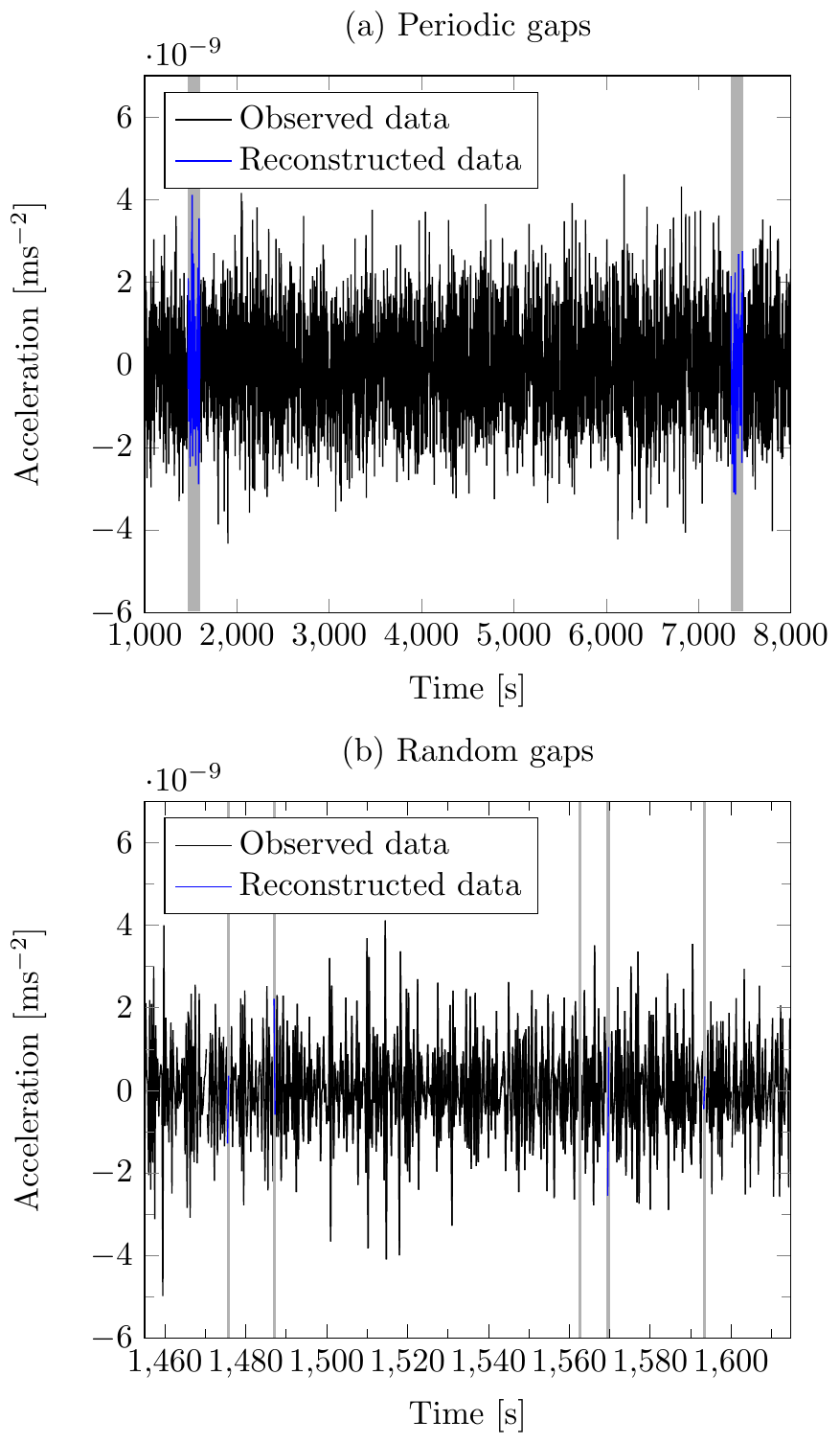}%
\caption{Extract of the time series obtained for a 20 orbit spin session sampled at 4 Hz, for the periodic gaps (top) and the random gaps (bottom). For clarity we plotted a sample which is longer for periodic gaps (lasting more than one orbital period $T_{\rm{orb}}$, or about 6 WEP periods $1/f_{\rm{EP}}$) than for random gaps (which are 260 times shorter and more scattered). Observed data are in black and reconstructed data from one Monte-Carlo conditional draw are in blue. The missing data spans are indicated by gray areas.}%
\label{fig:1}%
\end{figure}

To show the noise dominating the data, the gap distribution and the reconstruction of missing values, we plotted in Fig.~\ref{fig:1} an extract of the observed time series (in black) as well as the estimated missing data (in blue) in the case of the periodic and the random gaps. In the first case we see that the periodic gaps are concentrated whereas the random gaps are distributed throughout the time series. The pattern that we see on the top of Fig.~\ref{fig:1} for the periodic window is repeated every $T_{\rm{orb}}$.

\begin{figure*}%
\flushleft
\includegraphics[width=1.0\textwidth]{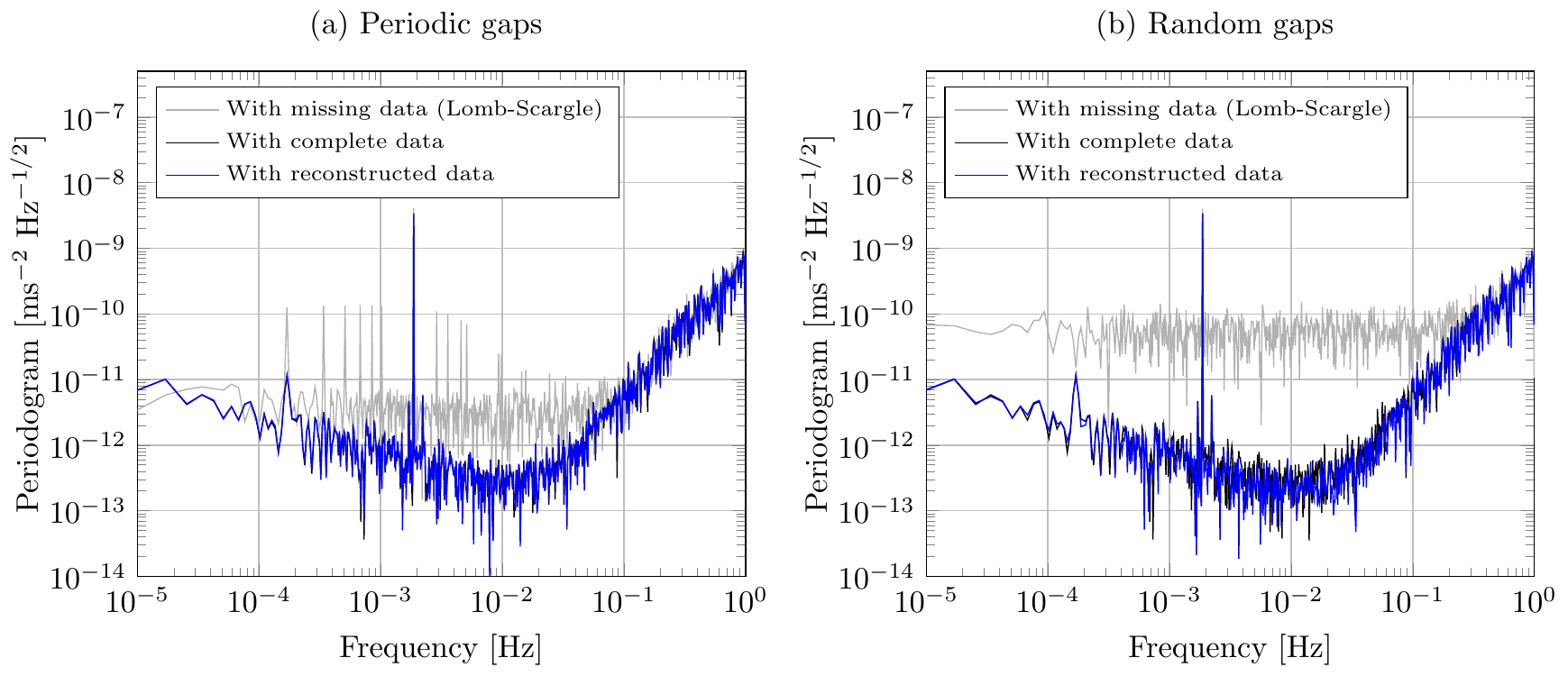}%
\caption{Lomb-scargle periodogram of the observed data (gray), periodogram of the reconstructed data (blue) and of the original data (black) for a 20-orbit spin session and for the periodic (left) and random (right) gap patterns. In order not to overload the figure the values have been plotted on a sub-grid of about 700 Fourier frequencies. The harmonic peaks of highest amplitudes visible around $2 f_{\rm{EP}} = 1.87 \times 10^{-3}$ Hz and common to the blue and black periodograms are not an artifact but are due to the gravitational gradient perturbation $T_{xx}$ and $T_{xz}$ (which are included in the model $A$). This is also the case for the peak at $f_{\rm{orb}}=1.70 \times 10^{-4}$ Hz.}%
\label{fig:2}%
\end{figure*}

In Fig.~\ref{fig:2} the Lomb-Scargle periodogam of the observed data is plotted without reconstruction (in light gray). This power spectrum estimate is adapted to unevenly spaced data and constructed so as to have similar statistical properties as the classical periodogram of Eq.~(\ref{eq:def_periodogram}) in the case of white noise. Then we plot the periodogram of the original complete data (black) and the periodogram of the reconstructed data (blue). 
We see that the reconstruction allows a more faithful visualization of the real noise level. Indeed with respect to the true original periodogram for complete data, the Lomb-Scargle periodogram exhibits spurious peaks in the case of periodic gaps while it shows a noise level which is higher by almost two orders of magnitude in the case of random gaps. The leakage level is not the same for random and periodic gaps, although the fraction of missing data is the same. This is due to the number of gaps as detailed in \cite{Berge2015}. In comparison the data reconstruction cancels the leakage effect that is present in the Lomb-Scargle periodogram when data are missing. We study the average behavior of the reconstructed periodogram in the next section.

\subsection{\label{subsec:PSD_results}Convergence of the periodogram and the PSD estimate}
\begin{figure*}%
\flushleft
\includegraphics[width=1.0\textwidth]{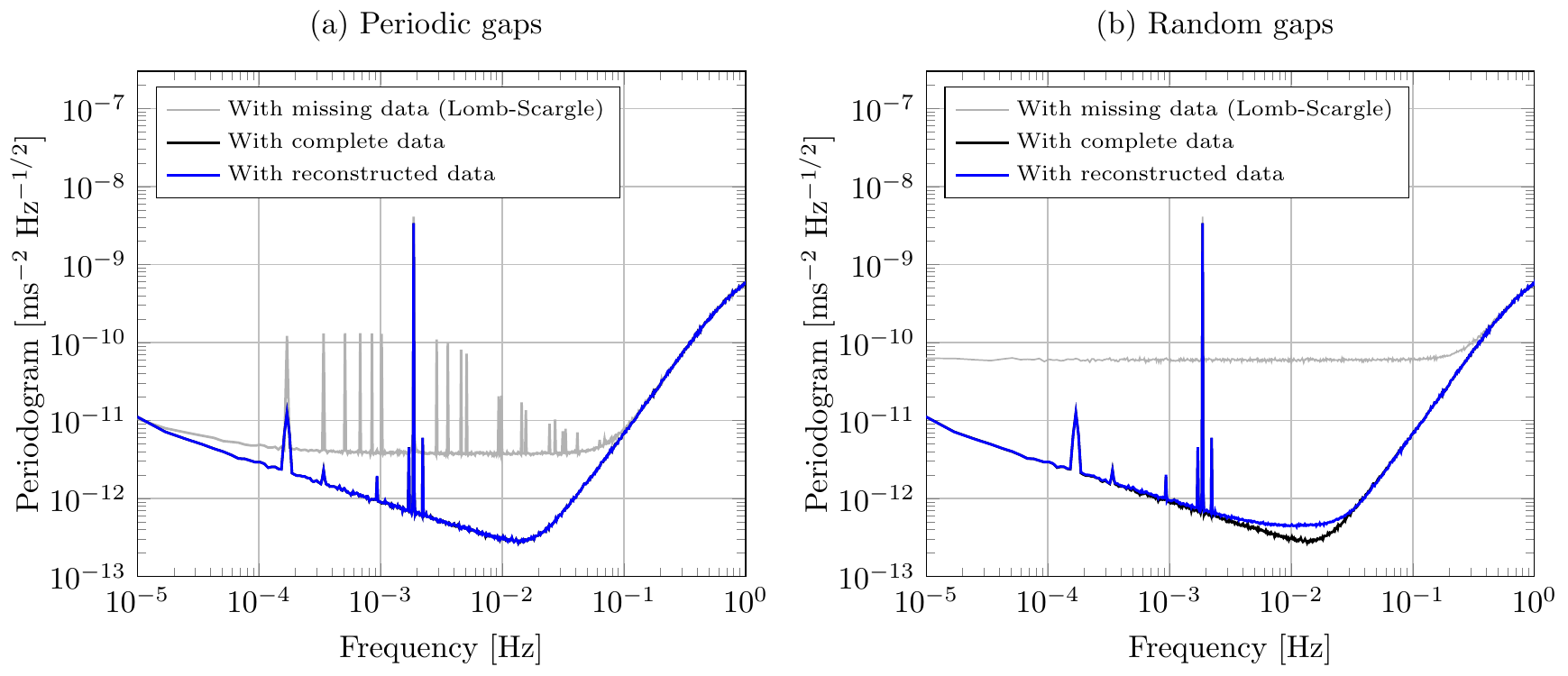}%
\caption{Average of 400 Lomb-scargle periodogram of the observed data (gray), of the periodogram of the reconstructed data (blue) and of the original data (black) for a 20-orbit spin session and for the periodic (left) and random (right) gap patterns. The same remarks made in Fig.~\ref{fig:2} about the peaks of the gravitational gradient perturbation apply. Averaging reveals the peak at $f_{\rm{EP}} = 9.35 \times 10^{-4}$ Hz that is due to the simulated WEP violation. It also makes another faint gradient harmonic visible at $2f_{\rm{orb}} = 3.40 \times 10^{-4}$ Hz which is due to the influence of the second zonal harmonic $J_2$.}%
\label{fig:3}%
\end{figure*}

Fig.~\ref{fig:3} shows the average of the periodograms over the 400 simulations: the Lomb-Scargle periodogram of the observed data (without any reconstruction), the complete data periodogram $I_y$ and the conditional periodogram $\mathrm{E}\left[ I_y |y_o \right]$. Note that they include the deterministic part of the signal since we have $\mathrm{E}\left[ I_y |y_o \right] = \mathrm{E}\left[ I_z |y_o \right] + I_{A\hat{\beta}}$.

Fig.~\ref{fig:3} shows that on average the conditional periodogram converges toward the true periodogram with no missing data. For random gaps however the conditional periodogram looks slightly biased between 3 and 30 mHz. This is mainly due to the low number of MC draws used to approximate the corrective term in Eq.~(\ref{eq:cond_periodogram}). To verify this explanation with a reasonable CPU time we have used a toy PSD model of almost similar shape with fewer data points and we have evidenced a decrease of the bias when increasing the number of MC draws to $N_d=100$ (see Appendix \ref{appendix:influence_MC_draws} for more details). Taking such a number of MC draws with $N = 4.7 \times 10^{5}$ data points would be computationally expensive but could be achieved by use of parallel computing. However this is not necessary for our purpose, as the signals of interest are located at lower frequencies than the biased interval. 

\begin{figure*}%
\flushleft
\includegraphics[width=0.95\textwidth]{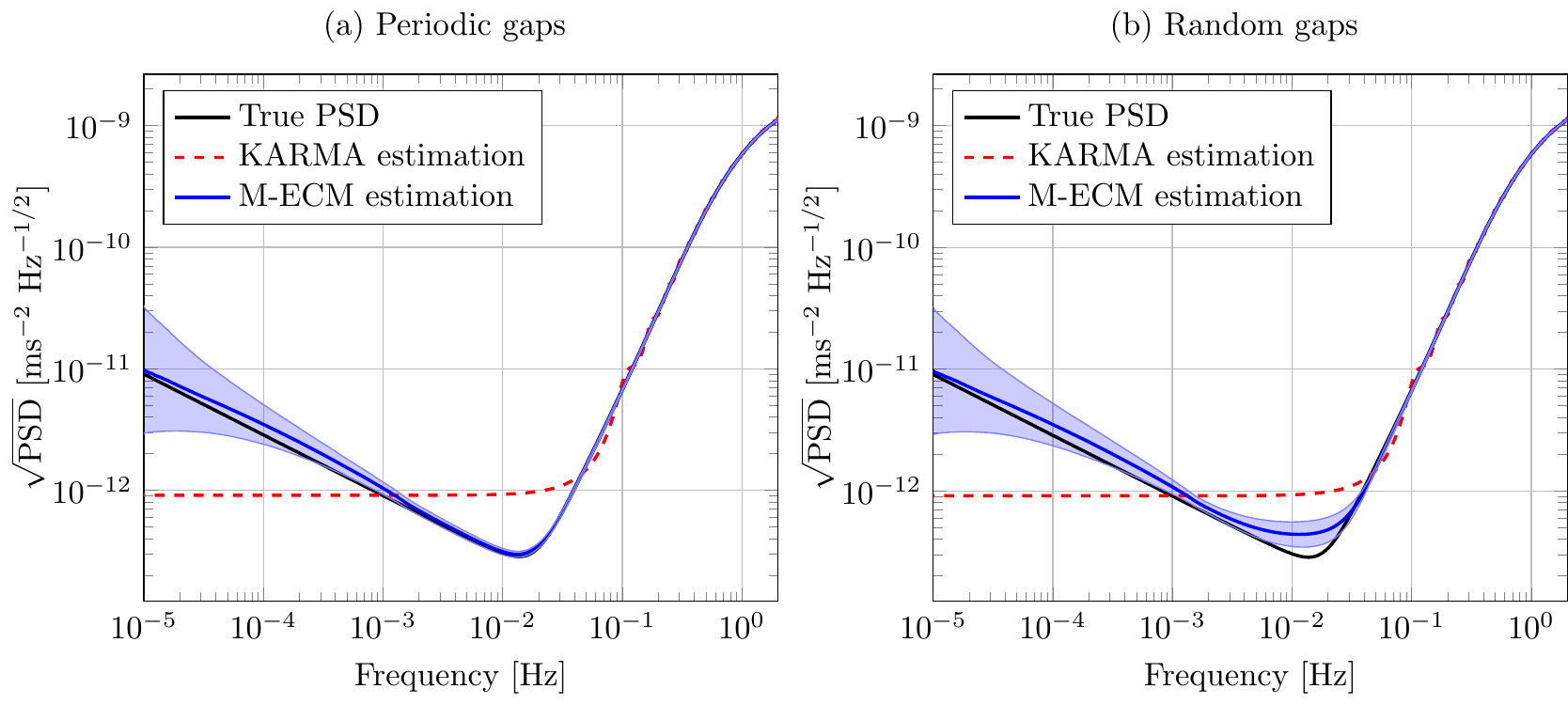}%
\caption{Sample average of the PSD estimates (blue) along with its 99\% confidence interval (light blue area), autoregressive estimate (dashed red) and true PSD (black) for a 20-orbit spin session and for the periodic (left) and random (right) gap patterns.}%
\label{fig:4}%
\end{figure*}

Fig.~\ref{fig:4} shows the average (blue) and the confidence level (light blue) of the PSD estimates. This average estimate is compared to the original PSD from which the noise is generated (black), 
and to the average estimate from the AR model of the KARMA method used as initial guess (dashed red).

The final PSD estimate brings an improvement with respect to the autoregressive one by reducing the bias in the low frequency part of the spectrum. 

In the case of random gaps, again a residual bias of less than $3 \times 10^{-13} \text{ } \rm{m}\rm{s}^{-2} \rm{Hz}^{-1/2}$ remains in the band between 3 and 30 mHz (where the PSD is minimum). Since the PSD is estimated from the periodogram, the bias in the periodogram whose origin is explained above has an impact on the bias of the PSD.

Another bias of order $10^{-12} \text{ } \rm{m}\rm{s}^{-2} \rm{Hz}^{-1/2}$ is still visible on both graphs (periodic and random gaps) at lower frequencies. This comes from another source of error: smoothing in the frequency domain. While reducing the variance, smoothing inevitably introduces a small bias visible even for the complete data case (we checked it by inspecting the PSD estimate obtained for complete data). A lead to slightly reduce this bias is to choose the optimal smoothing bandwidth $h_j$ as in \cite{FanGijbels1995} instead of the logarithmic function mentioned in Sec.~\ref{subsec:PSD_estimation}. The optimal bandwidth minimizes the Mean Squared Error (MSE) between the estimator and the true spectrum, but the minimization would be done at a higher computational cost since it requires to estimate the MSE on a grid of varying $h$ for each frequency.

\begin{figure}%
\flushleft
\includegraphics[width=1.0\columnwidth]{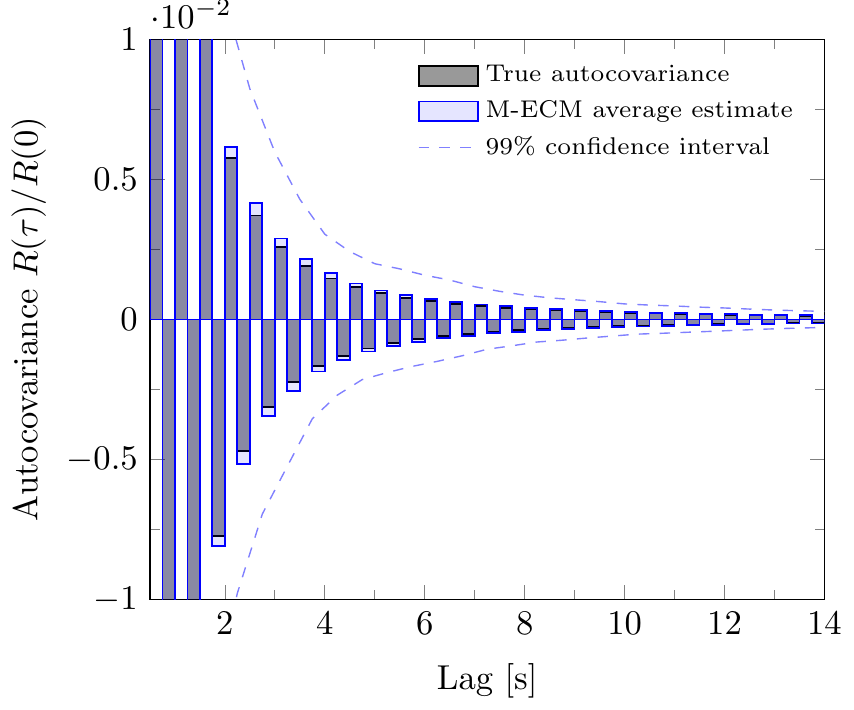}%
\caption{Sample average of the normalized autocovariance estimated from random-gapped data with the M-ECM algorithm (blue) along with its 99\% confidence interval (dashed blue) and the true autocovariance (black). The autocovariance alternates between positive and negative values. The bar heights of the histogram correspond to the values of $R(\tau)/R(0)$ at lags $\tau$ located at the left-hand-side edge of the bars. To be comprehensive the confidence interval should include both the upper and lower bounds for positive and negative values of the autocovariance. For the sake of clarity, only the upper bound of positive values and the lower bound of negative values are displayed. }%
\label{fig:5}%
\end{figure}

Finally, to give an insight of the short-range correlations of the noise and their estimation, we show in Fig.~\ref{fig:5} the estimate of the normalized autocovariance function $R(\tau)/R(0)$ obtained from the M-ECM PSD estimate with Eq.~(\ref{eq:autocov_PSD_estimate}), in the case of random gaps only since it corresponds to the largest error on the PSD. We can see an irregular pattern taking successively positive and negative values, which prevents us from fitting smooth autocovariance models commonly used in geostatistics as in \cite{Stroud}. According to the figure, although it is model-independent, the M-ECM algorithm provides an estimation of the covariance with a low bias.

\subsection{\label{subsec:regression_results}Performance of the regression parameters estimation}

Here we examine the empirical mean and variance of the regression vectors $\beta$ obtained with the M-ECM algorithm applied to the 400 simulated samples.

We gather in Table \ref{tab:regression_results} the sample average of the 400 estimates $\hat{\mu}_{400}$ of the regression parameter vector $\beta$, as well as their sample standard deviation $\hat{\sigma}_{400}$. 
Three estimation methods are compared: the ordinary least squares from the observed data described by Eq.~(\ref{OLS_estimator}), the KARMA method (Sec.~\ref{sec:KARMA}), and the M-ECM algorithm (Sec.~\ref{sec:refinement}). 
We check that the KARMA and M-ECM estimates both converge towards the true expectation value and therefore are unbiased. In addition the standard deviation of the M-ECM regression is comparable to KARMA's. A slight increase of uncertainty is observed for the M-ECM algorithm but this is not significant with respect to the precision of the sample standard deviation. Therefore the reconstruction is consistent and there is no loss of precision nor additional bias when performing a linear regression on data reconstructed with the M-ECM algorithm. Like the KARMA method the M-ECM algorithm allows us to improve the precision of the regression with respect to the OLS: the uncertainty is decreased by a factor 4 in the case of periodic gaps and by nearly a factor 60 in the case of random gaps. That is not because the M-ECM algorithm is less efficient with periodic gaps, but because the leakage effect is smaller with periodic gaps than with random gaps (see comment in Sec.~\ref{reconstruction_results}).

These results are compared to the minimal standard deviation that would be reached if the covariance were known exactly. This minimum is the Cram\'{e}r-Rao Lower Bound (CRLB) of the GLS estimator which is computed via Eq.~(\ref{eq:GLS_covariance}) by using the true covariance matrix derived from the true noise PSD $S(f)$. The CRLB is computed for each observation window and provides the reference to assess the performance of our algorithm. It is indicated in the third column of Table \ref{tab:regression_results}: both for the KARMA and M-ECM methods the estimation of $\delta$ has an uncertainty close to the CRLB within a difference less or equal to $10^{-16}$.

\begin{table*}
\caption{\label{tab:regression_results}Mean and standard deviations of the regression parameters with OLS, KARMA, and the M-ECM method. The first four columns indicate respectively the gap pattern, the parameter, its true value and the Cramer-Rao lower bound (minimal achievable standard deviation). Then from left to right there are 3 column groups corresponding to the 3 estimation methods. For all methods the quantities $\hat{\mu}_{400}$ and $\hat{\sigma}_{400}$ respectively correspond to the sample average and the sample standard deviation of the 400 estimates. For OLS the term $\sigma_{\rm{OLS}}$ is the theoretical error calculated with Eq.~(\ref{eq:covariance}), for the KARMA method $\hat{\sigma}_{\rm{AR}}$ is the average of the error estimates (see \cite{Baghi} for details) and for the M-ECM algorithm $\hat{\sigma}_{\rm{ECM}}$ is the average of the error estimates calculated with Eq.~(\ref{eq:GLS_covariance}).} 
\begin{center}
\begin{ruledtabular}
\begin{tabular}{ c  c  c  c | c  c  c | c  c  c | c  c  c }
\multirow{2}{*}{}                    & \multicolumn{3}{c|}{}             & \multicolumn{3}{c|}{Ordinary least squares}    & \multicolumn{3}{c|}{KARMA}  & \multicolumn{3}{c}{M-ECM} \\
    {Window}                         & {Param.}      & {True} & {CRLB} & {$\hat{\mu}_{400}$} & {$\sigma_{\rm{OLS}}$} & {$\hat{\sigma}_{400}$} & {$\hat{\mu}_{400}$} & {$\hat{\sigma}_{\rm{AR}}$} & {$\hat{\sigma}_{400}$} & {$\hat{\mu}_{400}$} & {$\hat{\sigma}_{\rm{ECM}}$} & {$\hat{\sigma}_{400}$} \\      \hline
			
    \multirow{3}{*}{Complete data}   & $\delta$  [$10^{-15}$] & $3$ & $0.96$ & $3.01$  & $1.08$ & $1.02$   & $2.98$ & $0.92$  & $0.96$  & $-$ & $-$ & $-$ \\
		                                 & $\Delta_{x}$ [$\mu$m] & $20$ & $0.003$ & $20.0$ & $0.005$ & $0.005$  & $20.0$ & $0.004$ & $0.003$ & $-$ & $-$ & $-$ \\
		  															 & $\Delta_{z}$ [$\mu$m] & $20$ & $0.003$ & $20.0$ & $0.005$ & $0.005$  & $20.0$ & $0.004$ & $0.003$ & $-$ & $-$ & $-$ \\
		
		\hline

		\multirow{3}{*}{Periodic gaps} & $\delta$  [$10^{-15}$] & $3$ & $0.97$ & $2.68$  & $4.13$ & $4.12$  & $2.98$     & $0.95$    & $0.97$  & $3.01$ & $1.11$ & $1.02$ \\ 
		                               & $\Delta_{x}$ [$\mu$m] & $20$  & $0.003$ & $20.0$ & $0.018$ & $0.018$  & $20.0$     & $0.004$   & $0.003$ & $20.0$ & $0.003$ & $0.005$ \\
		  													   & $\Delta_{z}$ [$\mu$m] & $20$ & $0.003$ & $20.0$  & $0.020$ & $0.019$  & $20.0$     & $0.005$   & $0.003$ & $20.0$ & $0.004$ & $0.005$ \\		
		
		\hline
		
    \multirow{3}{*}{Random gaps}   & $\delta$ [$10^{-15}$]  & $3$    & $1.05$& $8.82$ & $62.3$ & $65.2$  &  $2.98$    & $1.19$    & $1.14$  & $3.02$ & $1.12$ & $1.15$ \\
		                               & $\Delta_{x}$ [$\mu$m]  & $20$  & $0.004$ & $20.0$ & $0.290$ & $0.296$ &  $20.0$    & $0.006$  & $0.004$ & $20.0$ & $0.003$ & $0.007$ \\
		  														 & $\Delta_{z}$ [$\mu$m]  & $20$  & $0.004$ & $20.0$ & $0.290$ & $0.314$ &  $20.0$    & $0.006$  & $0.005$ & $20.0$ & $0.004$ & $0.007$ \\

\end{tabular}
\end{ruledtabular}
\end{center}
\end{table*}

\subsection{\label{subsec:precision_results}Results of the precision assessment}

We consider the results obtained for the M-ECM error estimate $\hat{\sigma}_{\rm{ECM}}$ defined in Sec.~\ref{subsec:error_assessment}, which is calculated for each data sample. Its sample average over the 400 simulations is shown in Table \ref{tab:regression_results}. They are in fair agreement with the sample standard deviations, which means that the M-ECM error estimate is a reliable way to assess the uncertainty of the regression from a single time series. In addition the variability of the M-ECM error estimate is low, being less than $5 \times 10^{-17}$ on average for $\hat{\delta}$ for both periodic and random gaps.

In addition it must be noted that the M-ECM error estimate is generally more reliable than the KARMA error estimate $\hat{\sigma}_{\rm{AR}}$ which tends to be biased by the error made in the estimation of the noise PSD at low frequencies (see Fig.~\ref{fig:4}). 
By better estimating the PSD with the M-ECM approach we increase the reliability of the error assessment made from a single measurement.

\section{\label{sec:conclusion}Conclusion}
We implement a Gaussian regression method valid in the context of gridded stationary Gaussian processes with missing data and unknown, arbitrary smooth PSD. 
A previously developed linear regression method based on an autoregressive noise model, efficiently estimating the regression parameters with nearly minimal variance, and providing a preliminary estimation of the noise PSD, is used as a basis to reconstruct the data in the missing intervals. This is done by conditional expectation of the deterministic and the stochastic parts of the missing data. To provide consistent reconstructed time series, the reconstruction and estimation steps are reproduced several times until convergence using an ECM-like algorithm. To adapt the ECM algorithm to spectral density estimation the PSD update step is performed using spectrum smoothing by a local formulation of the likelihood. This leads to maximizing the full likelihood by using a more general, model-independent description of the residuals in comparison to the autoregressive approach. 

We finally apply the developed tools to representative simulated data of the MICROSCOPE space mission and we demonstrate their accuracy and precision for two relevant patterns of gaps that can be encountered during the satellite flight, namely periodic gaps and stationary random gaps. Consistent with the autoregressive method, the results of the regression are unbiased and have a precision close to the minimal variance bound, which is respectively 4 to 60 times better than the ordinary least squares for the two gap patterns. We show that in the low frequency region of the spectrum the result of the PSD estimation is improved compared to the initial autoregressive guess, leading to a better characterization of the noise with an error less than $10^{-12}$ $\rm{m}\rm{s}^{-2} \rm{Hz}^{-1/2}$. In addition the modified ECM algorithm provides reconstructed data which are faithful to the true complete data. Indeed the reconstructed periodograms are on average consistent with the original ones within a few $10^{-13}$ $\rm{m}\rm{s}^{-2} \rm{Hz}^{-1/2}$. 

Besides, the developed ECM algorithm still leaves room for improvements particularly regarding the estimation of the PSD. The spectrum smoothing involved in the PSD estimation can be made optimal by use of automatic smoothing bandwidth selection. Such a refinement requires additional calculations whose cost has not been considered worth for our application (where the number of data points lies between $10^5$ to $10^6$) but may be acceptable for medium-size problems. We also point out that the conditional expectation of the periodogram can be more accurately estimated by increasing the number of Monte Carlo draws of the missing data, again increasing the computation time. These parameters can be adjusted depending on the level of desired accuracy. 

\appendix

\section{\label{appendix:cond_per}Calculation of the conditional expectation of the periodogram}
\label{annex:periodogram_cond}
Here we derive the expression of the conditional expectation of the periodogram of the noise $I_{z}$, which is defined by Eq.~(\ref{eq:def_periodogram}) taking $x$ to be the model residuals:
\begin{equation}
z = y - A\beta.
\label{eq:model_residuals}
\end{equation}
The periodogram of the noise calculated at the Fourier frequencies $f_k$ then read
\begin{equation}
I_z(k) = \frac{1}{N} \left| \sum_{n=0}^{N-1} z_{n} e^{-2 I \pi k n / N }\right|^{2}
\label{eq:periodogram_z_fourier}
\end{equation}

We first note that $I_z(k)$ is the diagonal element of the matrix $\tilde{z}\tilde{z}^{\dag}$ where $\tilde{z} = F_{N} z$ is the normalized discrete Fourier transform of the residual vector $z$. This vector can be decomposed into an observed part and a missing part:
\begin{equation}
z = W_{o}^{\dag} z_o + W_{m}^{\dag} z_m,
\label{eq:z_decomposition}
\end{equation}
where $W_{o}$ and $W_{m}$ are indicator matrices defined in Sec.~\ref{subsec:meas_eq}.
Thus we have:
\begin{equation*}
zz^{\dag} = W_{o}^{\dag} z_o z_o^{\dag} W_{o} + W_{m}^{\dag} z_m z_m^{\dag} W_{m} + W_{o}^{\dag} z_o z_m^{\dag} W_{m} + W_{m}^{\dag} z_m z_o^{\dag} W_{o}.
\label{eq:zz}
\end{equation*}
By taking the conditional expectation of this equation we get:
\begin{eqnarray}
\mathrm{E}\left[ zz^{\dag} |y_o \right] &=& W_{o}^{\dag} z_o z_o^{\dag} W_{o} + W_{m}^{\dag} \mathrm{E}\left[ z_m z_m^{\dag} |y_o \right]  W_{m} \nonumber \\ 
&& + W_{o}^{\dag} z_o \mu_{z_m|o}^{\dag} W_{m} + W_{m}^{\dag} \mu_{z_m|o} z_o^{\dag} W_{o},
\label{eq:cond_zz}
\end{eqnarray}
where we have set $\mu_{z_m|o} = \mathrm{E}\left[ z_m  |y_o \right]$.
By definition of the conditional covariance of the missing residuals, we have:
\begin{equation}
\mathrm{E}\left[ z_m z_m^{\dag} |y_o \right] = \mu_{z_m|o} \mu^{\dag}_{z_m|o} + \Sigma_{m|o}.
\label{eq:cond_exp_z_m_rearranged}
\end{equation}
Injecting this expression into Eq.~(\ref{eq:cond_zz}) we obtain:
\begin{eqnarray*}
\mathrm{E}\left[ zz^{\dag} |y_o \right] &=& W_{o}^{\dag} z_o z_o^{\dag} W_{o} + W_{m}^{\dag} \mu_{z_m|o} \mu_{z_m|o}^{\dag}  W_{m}  \\
&& + W_{m}^{\dag} \Sigma_{m|o} W_{m} + W_{o}^{\dag} z_o \mu_{z_m|o}^{\dag} W_{m} \\
&& + W_{m}^{\dag} \mu_{z_m|o} z_o^{\dag} W_{o}.
\end{eqnarray*}
To simplify this equation, we define the reconstructed residuals as
\begin{equation}
\hat{z} \equiv \mu_{z|o} = W_{o}^{\dag} z_o + W_{m}^{\dag} \mu_{z_m|o},
\label{eq:reconstructed_residuals}
\end{equation}
whose elements are equal to $z_i$ when $y_i$ is observed and to its conditional expectation $\mu_{z|o,i}$ when $y_i$ is missing. Using this definition we get:
\begin{equation}
\mathrm{E}\left[ zz^{\dag} |y_o \right] = \hat{z} \hat{z}^{\dag} + W_{m}^{\dag} \Sigma_{m|o} W_{m}.
\label{eq:cond_exp_zz_simplified}
\end{equation} 
Thus the conditional expectation of $\tilde{z}\tilde{z}^{\dag}$ is given by the matrix:
\begin{equation}
\mathrm{E}\left[ \tilde{z}\tilde{z}^{\dag} |y_o \right] = \tilde{\hat{z}} \tilde{\hat{z}}^{\dag} + F_{N} W_{m}^{\dag} \Sigma_{m|o} W_{m} F_{N}^{\dag}.
\label{eq:cond_exp_zz_Fourier}
\end{equation} 
The conditional expectation of the noise periodogram is finally given by the diagonal elements of the matrix \ref{eq:cond_exp_zz_Fourier}:
\begin{eqnarray}
\mathrm{E}\left[ I_z(k) |y_o \right] &=& \mathrm{E}\left[ \tilde{z}\tilde{z}^{\dag} |y_o \right]_{k,k} \nonumber \\ 
             &=& \frac{1}{N} \left| \sum_{n=0}^{N-1} \hat{z}_{n} e^{-2 I \pi k n / N }\right|^{2} + \sigma_k^{2},
\label{eq:cond_periodogram_k}
\end{eqnarray}
where we defined $\sigma_k^{2}$ as being the diagonal element of the matrix $F_{N} W_{m}^{\dag} \Sigma_{m|o} W_{m} F_{N}^{\dag}$.

\section{\label{appendix:cond_generation}Conditional generation of the periodogram}
The purpose of this appendix is to efficiently compute realizations of a random vector whose mean is the conditional expectation of the periodogram given by Eq.~(\ref{eq:cond_periodogram_k}). 
We consider the vector $z^{*}$ of size $N$ drawn from the distribution $\mathcal{N}(0,\Sigma)$. 
As mentioned in \cite{Dietrich}, the vector of size $N_m$ constructed as
\begin{equation}
z^{*}_{m|o} = \mu_{z|o} + W_{m} z^{*} - \Sigma_{mo} \Sigma_{oo}^{-1} W_{o} z^{*}
\label{eq:conditional_vector}
\end{equation}
has mean $\mu_{z_m|o}$ and covariance $\Sigma_{m|o}$.

We then construct the vector:
\begin{equation}
z^{*}|o = W_o^{\dag} z_o + W_m^{\dag} z^{*}_{m|o}
\label{eq:random_vector}
\end{equation}
We verify that this vector has the desired covariance as given by Eq.~(\ref{eq:cond_exp_zz_simplified}):
\begin{eqnarray}
\mathrm{E}\left[ (z^{*}|o) (z^{*}|o)^{\dag} \right] &=&  W_o^{\dag}  z_o z_o^{\dag} W_o+ W_m^{\dag} \mathrm{E}\left[z^{*}_{m|o}{z_{m|o}^{*}}^{\dag}\right] W_m \nonumber \\
&& + W_o^{\dag}  z_o \mu_{z_m|o} W_m + W_m^{\dag}  \mu_{z_m|o} z_o^{\dag} W_o \nonumber
\label{eq:verify}
\end{eqnarray}
By using the definition of the reconstructed residual vector $\hat{z}$ in Eq.~(\ref{eq:reconstructed_residuals}) and the definition of the covariance $\mathrm{E}\left[z^{*}_{m|o}{z_{m|o}^{*}}^{\dag}\right] = \mu_{z_m|o} + \Sigma_{m|o}$ we end up with the right-hand side of Eq.~(\ref{eq:cond_exp_zz_simplified}). Thus the periodogram of $z^{*}|o$ will have its expectation given by Eq.~(\ref{eq:cond_periodogram_k}), which is what we want. 

\section{\label{appendix:influence_MC_draws}Impact of the number of Monte-Carlo draws on the estimation of the conditional periodogram}
In Appendix \ref{appendix:cond_generation} we showed how to approximate the conditional expectation of the periodogram by multiple imputations (\textit{i.e.} Monte-Carlo draws) of the missing data. Here we check that the number of Monte-Carlo draws drives the accuracy of the conditional periodogram with respect to the original one.
In other words, the objective is to check that the mean of the conditional periodogram converges towards the mean of the original periodogram. Formally we verify the equality:
\begin{equation}
\mathrm{E}\left[ \mathrm{E}\left[I_y | y_o\right] \right] = \mathrm{E}\left[I_y\right].
\label{eq:equality_per_mean}
\end{equation}
To compute a reliable approximation of the expectations involved in the above equation we need to draw several realizations of the vector $y$. For each realization the conditional expectation $\mathrm{E}\left[I_y | y_o\right]$ requires to generate $N_d$ conditional draws of the missing data. We aim to check that increasing $N_d$ improves the ability of the left-hand side of Eq.(\ref{eq:equality_per_mean}) to approximate the right-hand-side. 

For the computational cost of this simulation to be acceptable we reduce the size of the problem to $N = 1000$ data points, and we modify the PSD model in Eq.~(\ref{eq:PSD_model}) to shift its minimum to higher frequency in order for it to be visible in the shortened observation bandwidth. In the following we label the PSD of this ``toy model'' $S'(f)$ and all related quantities are identified with a prime.

As explained in Sec.~\ref{subsec:periodogram_expectation} multiple imputation of missing data allows us to approximate the second term $\sigma^2$ in Eq.~(\ref{eq:cond_periodogram}). To amplify its relative amplitude with respect to the first term $I_{\hat{z}}$ we sharpen the ``hollow'' of the PSD shape by increasing the high frequency slope:
\begin{equation}
S'(f) = \alpha_{0}' + \alpha_{-1}' f^{-1} + \alpha_{5}' f^5
\label{eq:PSD_toy_model}.
\end{equation}

We choose a random gap window $w$ representing 10\% data losses - still with the idea of increasing the difference between the two terms of Eq.~(\ref{eq:cond_periodogram}). To simplify and focus on the PSD estimation we remove the deterministic part of the model (\textit{i.e.} $A = 0$, so that $y = z$). 

We generate 400 realizations of a noise vector $z'$ whose PSD is given by Eq.~(\ref{eq:PSD_toy_model}). Then we run the M-ECM algorithm for all realizations of the observed data $z_o'$ to obtain the conditional periodograms and the PSD estimates. We do it twice: for $N_d=5$ and for $N_d=100$. 

\begin{figure}[!h]%
\flushleft
\includegraphics[width=0.9\columnwidth]{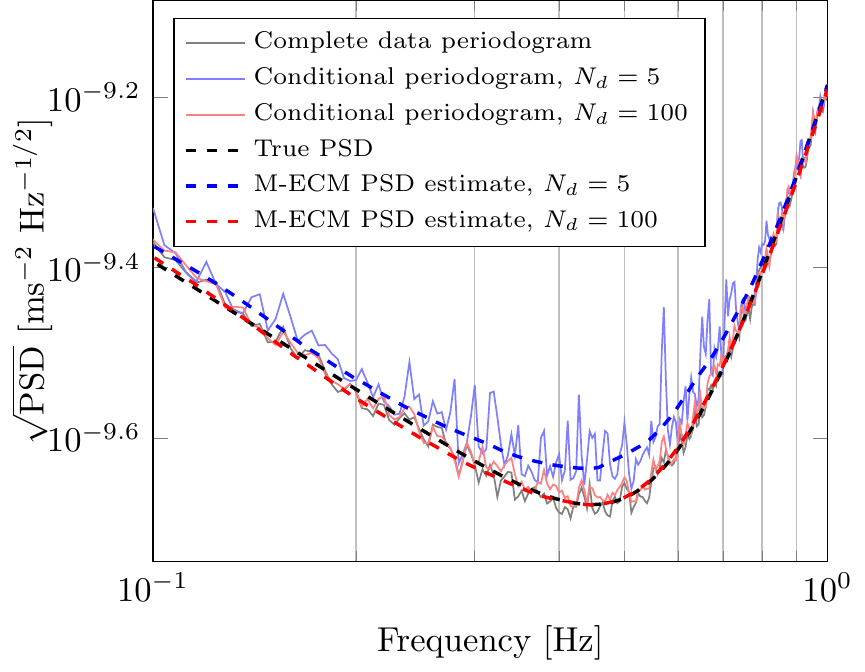}%
\caption{Average of 400 conditional periodograms and PSD estimates obtained with the M-ECM algorithm applied to data samples generated from the toy model defined by Eq.~(\ref{eq:PSD_toy_model}) with $N=1000$ data points. The blue and red curves correspond to numbers of MC draws respectively equal to $N_d=5$ and $N_d=100$. Light solid lines represent average periodograms while dashed lines represent PSDs. The gray curve is the average of the complete data periodogram and the black dashed curve is the true PSD. Therefore increasing $N_d$ lowers the bias of the PSD estimate.}%
\label{fig:6}%
\end{figure}

We finally calculate the sample average of the conditional periodograms and of the PSD estimates obtained with the two numbers of MC draws and show the result in Fig.~\ref{fig:6}. By comparing them to the true PSD we check that the mean of the estimate obtained with the largest $N_d$ is less biased than the other, which is what we meant to prove.

\FloatBarrier

\vspace{0.1cm}
\begin{acknowledgments}
The authors would like to thank all the members of the MICROSCOPE Performance team, as well as Jean Guerard and Sandrine Pires, for fruitful discussions. They would also like to thank the anonymous reviewer for his constructive review of their manuscript. This activity has been funded by ONERA and CNES. We also acknowledge the financial contribution of the UnivEarthS Labex program at Sorbonne Paris Cit\'e (ANR-10-LABX-0023 and ANR-11-IDEX-0005-02).
\end{acknowledgments}

\bibliography{references}

\end{document}